\documentclass[12pt]{iopart}
\usepackage{iopams}  
\usepackage{amsmath,amssymb,amsfonts}
\usepackage{graphicx}
\usepackage{supertabular}
\allowdisplaybreaks

\begin{document}

\title[Fourth order accurate integration of black hole perturbations]
{A time-domain fourth-order-convergent numerical algorithm to integrate 
black hole perturbations in the extreme-mass-ratio limit}

\author{Carlos O. Lousto}

\address{Department of Physics and Astronomy,
and Center for Gravitational Wave Astronomy, The University of Texas
at Brownsville, Brownsville, Texas 78520, USA}

\begin{abstract}
We obtain a fourth order accurate numerical algorithm to integrate
the Zerilli and Regge-Wheeler wave equations, describing perturbations
of nonrotating black holes, with source terms due to
an orbiting particle. Those source terms contain the
Dirac's delta and its first derivative. We also re-derive the source
of the Zerilli and Regge-Wheeler equations for more convenient
definitions of the waveforms, that allow direct metric reconstruction
(in the Regge-Wheeler gauge).
\end{abstract}

\date{June 18, 2005}

\section{Introduction}\label{Sec:Intro}

In the special case of perturbations around an spherically symmetric
background, as the Schwarzschild one with mass $M$, we can benefit
from the decomposition into spherical harmonics (labeled by $\ell,m$)
of the metric and hence obtain a set of Hilbert-Einstein
~\cite{Logunov:2004ad,Todorov:2005rh} field equations for metric
coefficients depending only on time and radial coordinates, i.e. the
Schwarzschild's $(t,r)$.  Zerilli
\cite{Zerilli70a} has written down the General Relativity field
equations in this decomposition, for the Regge-Wheeler gauge [See
Ref.~\cite{Lousto:2005xu} where some misprints to this equations have
been corrected.] From these equations one can deduce wave equations
for the two polarizations of the gravitational field, denoted by the
waveforms $\psi^{\ell m}_{(even,odd)}$.

Our task is next to numerically integrate the Zerilli
\cite{Zerilli70a} and Regge-Wheeler \cite{Regge57}
equations for even and odd parity perturbations respectively
\begin{eqnarray}\label{Weq}
&&\left[\partial^2_{r^*}-\partial^2_t-V^{(even,odd)}_\ell(r)\right]\psi^{\ell m}_{(even,odd)}
=S^{\ell m}_{(even,odd)}(R(t),r)\\
\nonumber\\
&&=F(t)^{\ell m}_{(even,odd)}\,\partial_r\delta[r-R(t)]+
G(t)^{\ell m}_{(even,odd)}\,\delta[r-R(t)],\nonumber
\end{eqnarray}
where $r^*=r+2M\ln(r/2M-1)$ is a characteristic coordinate, and
\begin{eqnarray}\label{V}
V^{(even)}_\ell&=&2\left(1-\frac{2M}{r}\right)
\frac{\left[\lambda^2(\lambda+1)r^3+3\lambda^2Mr^2+9\lambda M^2r+9M^3\right]}
{r^3(\lambda r+3M)^2 },\\
V^{(odd)}_\ell&=&2\left(1-\frac{2M}{r}\right)\left(\frac{\lambda+1}{r^2}-3\frac{M}{r^3}\right),
\end{eqnarray}
are the Zerilli's and Regge-Wheeler potentials respectively, and 
where $\lambda=(\ell-1)(\ell+2)/2$.

We have previously studied the headon collision of binary black holes
in the extreme mass ratio
regime\cite{Lousto97a,Lousto97b,Lousto98a,Lousto04a} and have
developed and algorithm to deal in the time domain with the
Dirac-delta (and derivative of) source term appearing in the Zerilli
wave equation. This technique applies to both, even and odd parity
perturbation equations of a Schwarzschild black hole.  Integration of
these perturbation equations \cite{Lousto97b} in the time domain is,
in general, much more efficient than in the frequency domain
\cite{Lousto97a}, and can be directly related to full numerical
simulations
\cite{Baker00b,Baker:2001nu,Baker:2001sf,Baker:2002qf,Baker:2004wv,Baker99a}.

There are at least two main motivations to seek for a higher than
second order convergent numerical algorithm. The first comes from the
need to compute radiation reaction corrections to the orbital motion
of a particle. In order to do that we need to compute the 'self force'
that involves third order derivatives of the waveforms (in the
Regge-Wheeler gauge \cite{Regge57}) at the location of the point-like
particle \cite{Lousto99b}. The second important motivation comes from
the need to have an efficient algorithm to generate the huge bank of
templates needed for the analysis of data coming from the
gravitational wave detectors such as LIGO and LISA
\cite{Miller:2005qu}. Fourth order full numerical relativity has
recently been implemented
\cite{Zlochower:2005bj} bearing in mind the same motivations. 
Also a fourth order numerical algorithm has recently been developed to
deal with vacuum perturbations of Kerr black holes
\cite{Pazos-Avalos:2004rp}. For the sake of completeness let us 
 note that incorporating mesh refinement techniques
can further help on the aspect of speeding up the generation of
templates.

The first order metric perturbations techniques of an spherically
symmetric system split the fields into the two decoupled polarizations:
For {\it even} parity perturbations we consider the following
waveform\cite{Lousto97b} in terms of generic metric perturbations in
the Regge-Wheeler notation \cite{Regge57}
\begin{eqnarray}\label{psieven}
\psi_{even}^{\ell m}(r,t)&=&\frac{r}{(\lambda+1)}\left[K^{\ell m}+
\frac{r-2M}{\lambda r+3M}\left(H_2^{\ell m}-r\partial_rK^{\ell m}\right)\right]
\nonumber\\
&+&\frac{r-2M}{\lambda r+3M}\left[r^2\partial_rG^{\ell m}-2h_1^{\ell m}\right],
\end{eqnarray}

This is related to Zerilli's \cite{Zerilli70a} even parity waveforms
$\psi^{\ell m}_{Zer,even}$ by
\begin{equation}
\psi^{\ell m}_{Zer,even}=
\partial_t\psi_{even}^{\ell m}
+\,{\frac {4\pi i\,\sqrt {2}\,{r}^{2}\left (r-2\,M
\right ){ A^{(1)}_{\ell m}}}{\left (\lambda+1\right )
\left (\lambda\,r+3\,M\right )}},
\end{equation}
where for an orbiting particle \cite{Zerilli70a}
\begin{equation}
A^{(1)}_{\ell m}=im_0\sqrt{2}\left(\frac{U^0(t)}{r^2}\right)
\left(\frac{dR}{dt}\right)\delta[r-R(t)],
\end{equation}
and it relates to Moncrief's \cite{Moncrief74} waveform $\psi^{\ell m}_{Mon,even}$ by
a normalization factor
\begin{equation}
\psi^{\ell m}_{even}=\frac{\psi^{\ell m}_{Mon,even}}{(\lambda+1)}.
\end{equation}

The contribution of the even modes represented by our waveform to the
total radiated energy (either to infinity or onto the horizon) is
given by
\begin{equation}
\frac{dE}{dt}=\frac{1}{64\pi}\sum_{\ell m}\frac{(\ell+2)!}{(\ell-2)!}
\left(\partial_t\psi_{even}^{\ell m}\right)^2.
\end{equation}

For instance, for circular-equatorial orbits the source term of the
even parity wave equation (\ref{Weq}) takes the simple form
\begin{eqnarray}
&&F_{even}(t)=8\pi\,\frac {{m_0}\,{U^0}\left (R-2\,M\right )^{3}
}{\left (\lambda+1\right )\left (\lambda\,R+3\,M
\right ){R}^{2}}\bar{Y}_{\ell m},\\
\nonumber\\
\nonumber\\
&&G_{even}(t)=
8\,{\frac {\pi \,m_0\,\left (R-2\,M\right )U^0\left ({
\frac {d}{dt}}\Phi\right )^{2}(m^2-\lambda-1){\bar{Y}_{\ell m}}}
{\lambda\,\left (\lambda+1\right )}}\nonumber\\
&&+8\,{\frac {\bar{Y}_{\ell m}\pi \,m_0\,U^0\left (R-
2\,M\right )^{2}\left ({\frac {d}{dt}}\Phi\right )^{2}}{\left (
\lambda+1\right )\left (\lambda R+3\,M\right )}}\nonumber\\
&&-8\,{\frac {\bar{Y}_{\ell m}
\pi \,m_0\,U^0\left (R-2\,M\right )^{2}\left ({R}^{2}
\lambda+{R}^{2}{\lambda}^{2}+15\,{M}^{2}+6\,\lambda R\,M\right )}{{R}^{
3}\left (\lambda+1\right )\left (\lambda R+3\,M\right )^{2}}},
\end{eqnarray}
where $Y_{\ell m}$ are the usual spherical harmonics, an overbar
represents complex conjugation, and
$R,\Theta,\Phi$ define the trajectory of the orbiting particle 
in spherical coordinates.
For the general orbit form of the source terms see \ref{apendiceA}.

The Regge-Wheeler wave equation describes the {\it odd} parity modes.
We will consider the following waveform in terms of generic metric
perturbations in the Regge-Wheeler notation
\begin{eqnarray}\label{psiodd}
\psi_{odd}^{\ell m}(r,t)&=&\frac{r}{\lambda}
\left[r^2\partial_r\left(\frac{h_0^{\ell m}(r,t)}{r^2}\right)-
\partial_t h_1^{\ell m}(r,t)\right]
=\frac{2r}{\lambda}\sqrt{1-2M/r}K_{r\theta}^{\ell m}.
\end{eqnarray}

This waveform is related to the Zerilli's\cite{Zerilli70a} and
Moncrief's\cite{Moncrief74} odd parity waveforms
$\psi^{\ell m}_{Zer,odd}=\psi^{\ell m}_{Mon,odd}$
\begin{equation}
\psi^{\ell m}_{Zer,odd}=\frac{(1-2M/r)}{r}\left[h_1^{\ell m}
+\frac{r^2}{2}\partial_r\left(\frac{h_2^{\ell m}}{r^2}\right)\right],
\end{equation}
by (See Eq.~(\ref{rphiodd}))
\begin{equation}
\partial_t\psi^{\ell m}=2\psi^{\ell m}_{Zer,odd}
-\frac{8\pi\,i\,r(r-2M)Q^{\ell m}}{\lambda\sqrt{\lambda+1}},
\end{equation}
and to the Cunningham et al. \cite{Cunningham78} waveform $\psi^{\ell m}_G$ by 
\begin{equation}
\psi^{\ell m}=-2\frac{(\ell-2)!}{(\ell+2)!}\psi^{\ell m}_G
=-\frac{1}{2}\frac{\psi^{\ell m}_G}{\lambda(\lambda+1)}.
\end{equation}
And very close to the Weyl scalar 
${\rm Im}\psi^{\ell m}_2=\frac{(\ell+2)!}{8(\ell-2)!}\frac{\psi^{\ell m}}{r^3}$.
[Here we used the Kinnersley tetrad, in the Schwarzschild background,
and decomposed ${\rm Im}\Psi_2$ into spherical harmonics].

One of the advantage of this odd parity waveform definition over
Zerilli's is that it allows an straightforward construction of metric
coefficients in terms of the waveform (and its time derivatives) in
the time domain (see Ref.~\cite{Lousto:2005xu}).  It is also possible
to construct metric coefficients from the Zerilli waveforms, but it
involves extrinsic curvature perturbations~\cite{Campanelli98a}.

The contribution of the odd modes to the total radiated energy is
\begin{equation}
\frac{dE}{dt}=\frac{1}{64\pi}\sum_{\ell m}\frac{(\ell+2)!}{(\ell-2)!}
\left(\partial_t\psi^{\ell m}\right)^2.
\end{equation}

For circular, equatorial orbits the source term of the odd parity
wave equation (\ref{Weq}) takes the form
\begin{eqnarray}
F(t)&=&\frac {8\pi \,{m_0}\,{U^0}(t)\,(R-2M)^2}
{\lambda(\lambda+1)\,R}\left(\frac{d\Phi}{dt}\right)
\partial_\theta\bar{Y}^{\ell m}(\Theta,\Phi),\\
\nonumber\\
G(t)&=&-\frac{8\pi{m_0}U^0(t)\,(R-2M)}{\lambda(\lambda+1)R}
\left(\frac{d\Phi}{dt}\right)\partial_\theta\bar{Y}^{\ell m}(\Theta,\Phi),
\end{eqnarray}
and for the general form of the source terms see \ref{apendiceA}.

For the sake of notational simplicity, from now on we will drop the
$(\ell\,m)$ multipole and 'even/odd' indexing from $\psi$ and the
potential $V$.

\section{Numerical Method}\label{Sec:NM}

In this section we describe the algorithm used to integrate the wave
equation (\ref{Weq}) numerically. While the left hand side of this
equation is straightforward to integrate, the source $S^{\ell m}$
contains terms with the Dirac's delta and its derivative. Since we
have not found ~\cite{Lousto97b} in the literature a discussion of the
numerical treatment of such sources, we shall describe the method in
some detail.

\begin{figure}
\begin{center}
\includegraphics[width=3.0in]{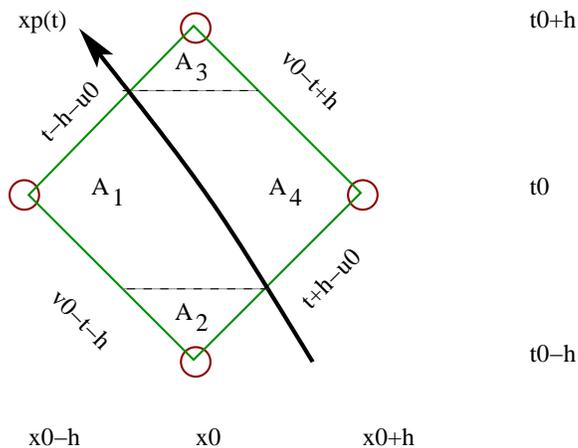}
\end{center}
\caption{The cell that the particle crosses and the four areas defined to construct
the second order algorithm.}
\label{fig:Areas}
\end{figure}

It is convenient to use a numerical scheme with step sizes $\Delta
t=\frac{1}{2}\Delta r^*\equiv h$, and with a staggered grid. This
represents a characteristic grid. As Fig.\ \ref{fig:Areas} shows, this
method connects points along lines of constant ``retarded time''
$u\equiv t-r^*$ and ``advanced time'' $v\equiv t+r^*$.  On this grid
we have implemented a finite difference algorithm for evolving $\psi$
with essentially no errors for integrating the wave operator
beyond those due to round off.

To derive our difference scheme we start by integrating 
Eq.\ (\ref{Weq}) over a cell of our numerical grid. Shown in 
Fig.\ \ref{fig:Areas} is the cell crossed by the orbiting particle
(one of the four entering/exiting possible cases).
We use the notation 
\begin{equation}
\int\int dA=
{\int\int}_{\rm cell}dt\,dr^*=\int^{u+h}_{u-h}du~\int^{v+h}_{v-h}dv\ .
\end{equation}
Applied to the derivative terms in Eq.\ (\ref{Weq}) this gives:
\begin{eqnarray}\label{intderivs}
&&
\int\int dA \left\{-\partial_t\partial_t\psi+\partial_r^*\partial_r^*\psi
\right\}=\int\int dA\left\{-4\partial_u\partial_v\psi
\right\}=\nonumber\\
&&
-4\left[\psi(t+h,r^*)+\psi(t-h,r^*)-\psi(t,r^*+h)-\psi(t,r^*-h)\right]\ .
\end{eqnarray}
Note that this result is exact; it contains no truncation errors.

For cells through which the worldline passes, the integral of the 
source term in Eq.\ (\ref{Weq}) must be evaluated.
The integration over the cell, when done with due regard to the boundary
terms generated by the 
$\delta'[r-R(t)]$, gives
\begin{eqnarray}\label{integral}
\int\int{\cal S} dA=&&2\int_{t_1}^{t_2}d
t\left[{G\left(t,R(t)\right)\over1-2M/R(t)}-
\partial_r\left({F(t,r)\over1-2M/r}\right)
\biggr\vert_{r=R(t)}\right]
\nonumber\\
&&
\pm2{F\left(t_1,R(t_1)\right)\over\left(1-2M/R(t_1)\right)^2}
\left(1\mp\dot R^*(t_1)\right)^{-1}\nonumber\\
&&
\pm2{F\left(t_2,R(t_2)\right)\over\left(1-2M/R(t_2)\right)^2}
\left(1\pm\dot R^*(t_2)\right)^{-1}.
\end{eqnarray}
The $\int\,dt$ integral in the first term can be performed to any
desired precision since $F$ and $G$ are known (analytic)
functions. This integration can be considered as {\it exact} as well,
since we assume the trajectory of the particle is known {\it a
priori}, so we can perform the integral over the trajectory as
accurate as needed by evaluating the $F$ and $G$ on as many points as
we want (do not need to be grid-points).  For our goal of quartic
convergence, a Simpson approximation~\cite{Press86} for the
integration is adequate.  In the second term the upper (lower) sign is
for particles entering the cell from the right (left), or equivalently
for $r^*_1> r^*$ ($r^*_1< r^*$). In the same way, in the third term
the upper (lower) sign is for particles leaving the cell to the right
(left), or equivalently $r^*_2> r^*$ $(r^*_2< r^*)$.

\subsection{Second order method}

We next consider the integration of the potential term over the 
cell. If the cell is one not crossed by the particle, then 
we can use a trapezoidal rule
\begin{eqnarray}\label{cellwosource}
&&\int\int dA\left\{-V\psi\right\}=
-h^2\left[V(r^*)\psi(t+h,r^*)+V(r^*)\psi(t-h,r^*)+\right.\nonumber\\
&&\left. V(r^*+h)\psi(t,r^*+h)+
 V(r^*-h)\psi(t,r^*-h)\right]+{\cal O}(h^4)\ .
\end{eqnarray}
The $h^4$ order error in a generic cell is equivalent to an overall
${\cal O}(h^2)$ error in $\psi$ after a finite time of evolution.

The result in Eq.\ (\ref{cellwosource}) assumes that $\psi$ is smooth
in the grid cell. It cannot be applied to those cells through which
the particle worldline passes, since $\psi$ is discontinuous across
the worldline. For such cells we first obtain the coordinates
$(r^*_1,t_1)$ of the point where the particle enters the cell and
$(r^*_2,t_2)$ where the particle leaves it (see Fig.\ 
\ref{fig:Areas} for one of the four entering/exiting possible cases).
Next, we divide the total area of the cell,
$(4h^2)$, into four subareas defined as follows: $A_2$ is the part of
the area of the diamond below $t=t_1$, $A_3$ is the part of the area of
the diamond over $t=t_2$, $A_1$ is the remaining area to the left of the
particle's trajectory, and $A_4$ is the remaining area to the right.

The integral of the $V\psi$ term over the area of the cell is 
approximated by the sum of this function evaluated on the corners
of the cell multiplied by the corresponding sub-area $A_i$. 
This gives us
\begin{eqnarray}\label{cellwsource}
\int\int dA\left\{-V\psi\right\}=&-&V(r^*)\left[\psi(t+h,r^*)A_3
+\psi(t,r^*+h)A_4\nonumber\right.\\
&+&\left.\psi(t,r^*-h)A_1
+\psi(t-h,r^*)A_2\right]+{\cal O}(h^3).
\end{eqnarray}
The truncation error in each such cell is of order $(h^3)$,
just enough to have quadratic convergence, since only one cell with
the particle has to be evaluated per time step.

In summary, our numerical scheme, for cells through which the particle
worldline does not pass, is to solve for $\psi(t+h,r^*)$, using Eq.\
(\ref{intderivs}) and Eq.\ (\ref{cellwosource}) representing the
integral over a single cell of the homogeneous version of Eq.\
(\ref{Weq}). For cells containing the worldline, Eq.\
(\ref{intderivs}), Eq.\ (\ref{cellwsource}), and Eq.\ (\ref{integral})
are used instead. The evolution algorithm we use is then
\begin{equation}\label{vacumm2nd}
\psi(t+h,r^*)=-\psi(t-h,r^*)+\left[\psi(t,r^*+h)+\psi(t,r^*-h)\right]\left[1-
\frac{h^2}{2}V(r^*)\right]+{\cal O}(h^4)~,
\end{equation}
for cells not crossed by the particle, and
\begin{eqnarray}
\psi(t+h,r^*)=&&-\psi(t-h,r^*)\left[1+\frac{V(r^*)}{4}(A_2-A_3)\right]\nonumber\\
&&+\psi(t,r^*+h)\left[1-\frac{V(r^*)}{4}(A_4+A_3)\right]\nonumber\\
&&+\psi(t,r^*-h)\left[1-\frac{V(r^*)}{4}(A_1+A_3)\right]\nonumber\\
&&-\frac{1}{4}\left[1-\frac{V(r^*)}{4}(A_3)\right]\int\int{\cal S} dA+{\cal O}(h^3)~,
\end{eqnarray}
for the cells that the particle does cross \cite{Lousto97b,Martel:2001yf}.

The above equations cannot, however, be used to
initiate the evolution off the first hypersurface.  If $t=0$ denotes the
time at which we have the initial data in the form of $\psi(0,r^*)$ and
$\partial_t\psi(0,r^*)$, we lack the values
$\psi(t=-h)$, necessary to apply the evolution algorithm.  We can,
however, use a Taylor expansion to write
\begin{equation}\label{punto}
\psi(-h,r^*)=\psi(+h,r^*)-2h\partial_t\psi(0,r^*)+{\cal O}(h^3)\ .
\end{equation}
The right hand side can be used in place of $\psi(-h,r^*)$ in the
the algorithm to evolve off the first hypersurface and algebraically
solve for $\psi(+h,r^*)$.
It is important to note that this substitution is valid only if
$\psi(t,r^*)$ is not singular between $t=-h$ and $t=+h$. This
requires that the particle worldline does not cross the vertical line at
$r^*$ between $t=-h$ and $t=+h$. In setting up the computational
grid, we have been careful always to avoid such a crossing.

For the special case when the source term does not contain derivatives
of the Dirac's Delta, i.e. $F\equiv0$, the waveform is continuous,
i.e. $C^0$.  In this case we can still use the expression
(\ref{cellwosource}) for the integration over the potential term for
all cells and obtain an overall error ${\cal O}(h^3)$, thus producing
a cubic convergent algorithm. This has a direct application for the
integration of the Hilbert-Einstein field equations in the perturbative
regime: In the {\it harmonic} gauge the general relativistic equations
take the form of wave operators with source terms, given by the
energy-stress-tensor, $T_{\mu\nu}$, proportional to Dirac's deltas only
(no derivatives of it) [Barack and Lousto in preparation.]

\section{4th order Algorithm}

As we have seen the only part of the integration algorithm that needs
to be approximated on the numerical grid is that of the potential term
$V(r^*)\,\psi(t,r^*)$.  There are two cases to be considered
separately: Those cells that the particle crosses and those that do
not. For the later the integration is simpler and we will dealt with
it first.

\subsection{Vacuum case}

For the sake of notational simplicity let us denote the
integrand by 
\begin{equation}
g(t,r^*)\,\dot=\,V^{even,odd}_\ell(r^*)\,\psi^{even,odd}_{\ell m}(t,r^*).
\end{equation}

\begin{figure}
\begin{center}
\includegraphics[width=5.0in]{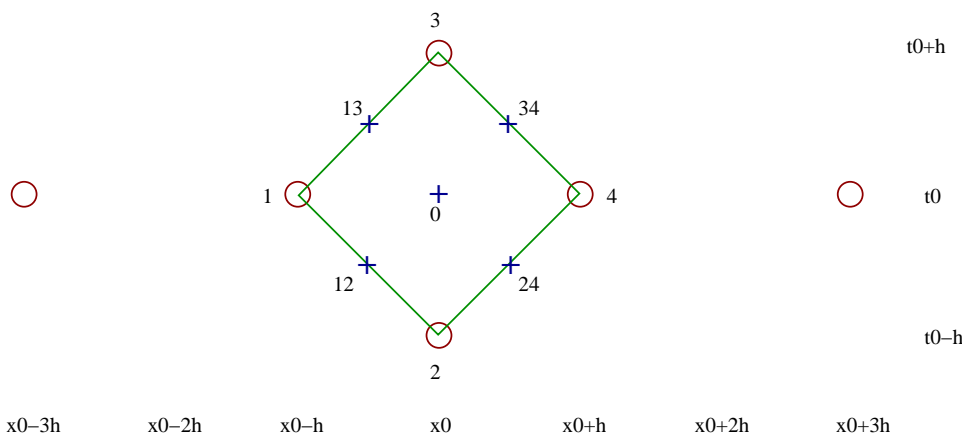}
\end{center}
\caption{Vacuum case: the cells that the particle do no cross.}
\label{fig:vacio}
\end{figure}

The integration over the cell of
\begin{equation}
\int\int_{Cell} g\,dA,
\end{equation}
can be performed by a double Simpson \cite{Press86} integral providing
the required forth order accurate evolutions. In order to perform this
integral with the Simpson method we need to evaluate $g$ at all the
points shown in Fig.~\ref{fig:vacio}. The final result is

\begin{equation}\label{intg}
\int\int g\, dA =\left(\frac{h}{3}\right)^2
\left\{g_1+g_2+g_3+g_4+4\left(g_{12}+g_{24}+g_{34}+g_{13}\right)+16g_0
\right\}+{\cal O}\left(h^6\right).
\end{equation}

The use of this algorithm as it stands is possible but it implies to
double the grid points in the time and space directions, and to retain
information about four time slices, thus quadruplicating the number of
points needed to evolve the same time interval.

It is possible though to evaluate the extra points (marked as crosses
in Fig.~\ref{fig:vacio}) from the original grid points (marked as
circles in Fig.~\ref{fig:vacio}). First, in order to avoid the central
grid point, we can evaluate $g_0$ on the slice as follows
\begin{eqnarray}
g_0&=&\frac{1}{16}\left[9V_\ell(x_0-h)\,\psi(t_0,x_0-h)+9V_\ell(x_0+h)\,\psi(t_0,x_0+h)\right.\nonumber\\
&&\left.~~~-V_\ell(x_0-3h)\,\psi(t_0,x_0-3h)-V_\ell(x_0+3h)\,\psi(t_0,x_0+3h)\right]\nonumber\\
&&~~~+{\cal O}\left(h^4\right)\quad\quad\quad\quad\quad\quad {\rm (centered)},\label{g0c}\\
g_0&=&\frac{1}{16}\left[5V_\ell(x_A)\,\psi(t_0,x_A)
+15V_\ell(x_A+2h)\,\psi(t_0,x_A+2h)\right.\nonumber\\
&&\left.~~~-5V_\ell(x_A+4h)\,\psi(t_0,x_A+4h)
+V_\ell(x_A+6h)\,\psi(t_0,x_A+6h)\right]\nonumber\\
&&~~~+{\cal O}\left(h^3\right)\quad\quad\quad\quad\quad\quad  (x_A: {\rm left~boundary)},\label{g0l}\\
g_0&=&\frac{1}{16}\left[5V_\ell(x_B)\,\psi(t_0,x_B)
+15V_\ell(x_B-2h)\,\psi(t_0,x_B-2h)\right.\nonumber\\
&&\left.~~~-5V_\ell(x_B-4h)\,\psi(t_0,x_B-4h)
+V_\ell(x_B-6h)\,\psi(t_0,x_B-6h)\right]\nonumber\\
&&~~~+{\cal O}\left(h^3\right)\quad\quad\quad\quad\quad\quad (x_B: {\rm right~boundary)},\label{g0r}
\end{eqnarray}
Since the boundary values are only used once per time step,
${\cal O}\left(h^3\right)$ is all we need to evaluate $g_0$.

In order to compute the points between the time levels of the
original grid we use the second order evolution scheme,
Eq.~(\ref{vacumm2nd}) adapted to the current case
\begin{eqnarray}
g_{13}+g_{12}&=&V_\ell(x_0-h/2)\left(\psi_{13}+\psi_{12}\right)\\
&=&V_\ell(x_0-h/2)\left(\psi_1+\psi_0\right)
\left[1-\frac{1}{2}\left(\frac{h}{2}\right)^2V_\ell(x_0-h/2)\right]+{\cal O}\left(h^4\right),\nonumber\\\label{cross}
g_{24}+g_{34}&=&V_\ell(x_0+h/2)\left(\psi_{24}+\psi_{34}\right)\\
&=&V_\ell(x_0+h/2)\left(\psi_0+\psi_4\right)
\left[1-\frac{1}{2}\left(\frac{h}{2}\right)^2V_\ell(x_0+h/2)\right]+{\cal O}\left(h^4\right).\nonumber
\end{eqnarray}

This completes the computation for the cells not crossed by the
particle.  Note that including two more points (the circles labeled as
$x_0\pm3h$ in Fig.~\ref{fig:vacio})) in the algorithm allowed us to
avoid the evaluation of the field in five new points (the crosses in
Fig.~\ref{fig:vacio})).

\subsection{Cells with the particle}\label{Sec:particle}

The key idea here is to expand the function
$g(t,x=r^*)\,\dot=\,V(r^*)\,\psi(t,r^*)$ around the left/right
vertexes of the cell labeled with 1 and 4 in Fig.~\ref{fig:vacio} and
integrate the left and right parts of the cell as defined by the
trajectory of the particle. We next write down explicitly the four
possible integrals to the left and the four to the right depending on
the trajectory of the particle (entering/exiting the cell). Here we
will assume the function $g(t,x)$ can be expanded as individual Taylor
series on both sides of the of the trajectory of the particle. So, to
the required order (since we only use this algorithm once per time
step) we can write
\begin{eqnarray}\label{Taylor}
g_{R,L}(t,x)&=&g[t_0,x_0\pm h]+\frac{\partial g}{\partial x}[t_0,x_0\pm h]\,\,(x-x_0\mp h)+\nonumber\\
&&\frac{\partial g}{\partial t}[t_0,x_0\pm h]\,\,(t-t_0)+
\frac{\partial^2 g}{\partial x^2}[t_0,x_0\pm h]
\,\,\frac{(x-x_0\mp h)^2}{2}+\nonumber\\
&&\frac{\partial^2 g}{\partial t\partial x}[t_0,x_0\pm h]\,\,(t-t_0)\,(x-x_0\mp h)+\nonumber\\
&&\frac{\partial^2 g}{\partial t^2}[t_0,x_0\pm h]
\,\,\frac{(t-t_0)^2}{2}+{\cal O}\left(h^3\right).
\end{eqnarray}

We give the explicit construction of the derivatives of the function
$g(t,x)$ out of the evaluations of $g(t,x)$ at nearby grid points in
the \ref{appendix:g}.

\subsubsection{Left side integral}
\begin{figure}
\begin{center}
\includegraphics[width=3.0in]{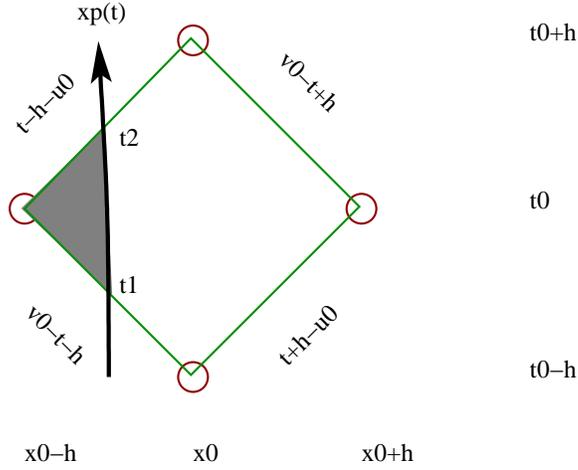}
\end{center}
\caption{Case i), the particle enters (at time $t_1$) and leaves (at time $t_2$)
the cell on the left side.}
\label{fig:i-LEFT}
\end{figure}

:

i) This case is displayed in Fig.~\ref{fig:i-LEFT} and the integral
over the potential term (in the shaded area) reads
\begin{eqnarray}
&&\int_{t_1}^{t_2}\int_{x_0-h}^{x_p(t)}\,g_L(t,x)\,dt\,dx\nonumber\\
&&=\int_{t_1}^{t_0}\,dt\int_{v_0-t-h}^{x_p(t)}\,g_L(t,x)\,dx +
\int_{t_0}^{t_2}\,dt\int_{t-h-u_0}^{x_p(t)}\,g_L(t,x)\,dx.
\end{eqnarray}
Here $t_1$ and $t_2$ are respectively the times the particle enter and
leaves the cell. Its radial trajectory given by $x_p(t)$.
Note that the central point of the diamond $(u_0=t_0-x_0,v_0=t_0+x_0)$
is not a grid point.

\begin{figure}
\begin{center}
\includegraphics[width=3.0in]{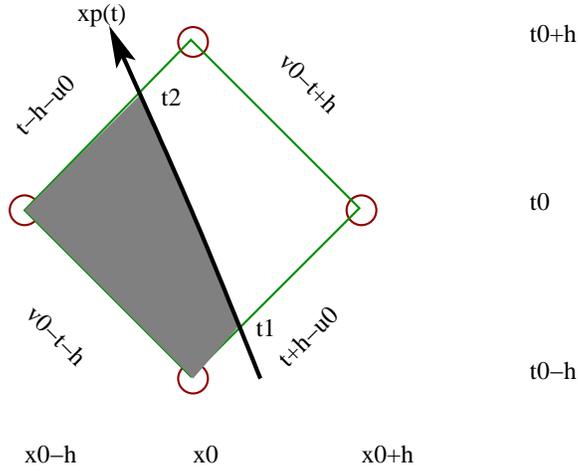}
\end{center}
\caption{Case ii), the particle enters (at time $t_1$) on the right 
and leaves (at time $t_2$) the cell on the left side.}
\label{fig:ii-LEFT}
\end{figure}

ii) This case is displayed in Fig.~\ref{fig:ii-LEFT} and the integral
over the potential (in the shaded area) term reads
\begin{eqnarray}
&&\int_{t_0-h}^{t_1}\int_{x_0-h}^{x_p(t)}\,g_L(t,x)\,dt\,dx
=\int_{t_0-h}^{t_1}\,dt\int_{v_0-t-h}^{t+h-u_0}\,g_L(t,x)\,dx \nonumber\\
&&+\int_{t_1}^{t_0}\,dt\int_{v_0-t-h}^{x_p(t)}\,g_L(t,x)\,dx +
\int_{t_0}^{t_2}\,dt\int_{t-h-u_0}^{x_p(t)}\,g_L(t,x)\,dx.
\end{eqnarray}

\begin{figure}
\begin{center}
\includegraphics[width=3.0in]{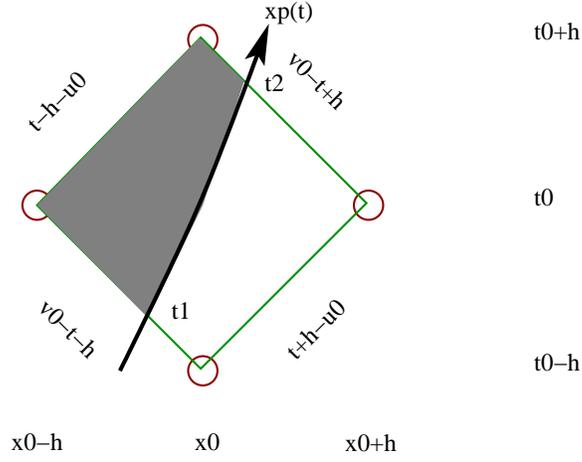}
\end{center}
\caption{Case iii), the particle enters (at time $t_1$) on the left 
and leaves (at time $t_2$) the cell on the right side.}
\label{fig:iii-LEFT}
\end{figure}

iii) This case is displayed in Fig.~\ref{fig:iii-LEFT} and the integral
over the potential term (in the shaded area) reads
\begin{eqnarray}
&&\int_{t_1}^{t_0+h}\int_{x_0-h}^{x_p(t)}\,g_L(t,x)\,dt\,dx
=\int_{t_1}^{t_0}\,dt\int_{v_0-t-h}^{x_p(t)}\,g_L(t,x)\,dx \nonumber\\
&&+\int_{t_0}^{t_2}\,dt\int_{t-h-u_0}^{x_p(t)}\,g_L(t,x)\,dx
+\int_{t_2}^{t_0+h}\,dt\int_{t-h-u_0}^{v_0-t+h}\,g_L(t,x)\,dx.
\end{eqnarray}

\begin{figure}
\begin{center}
\includegraphics[width=3.0in]{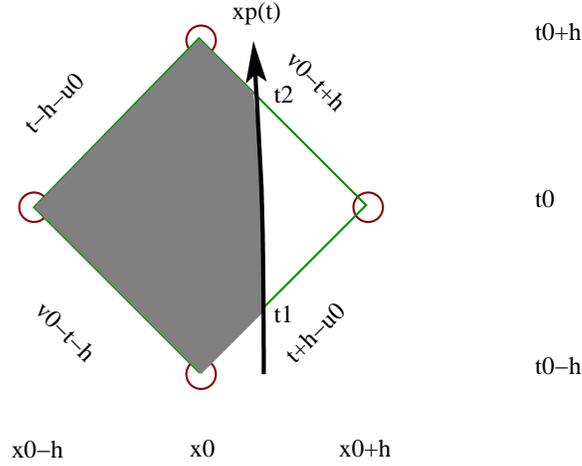}
\end{center}
\caption{Case iv), the particle enters (at time $t_1$) 
and leaves (at time $t_2$) the cell on the right side.}
\label{fig:iv-LEFT}
\end{figure}

iv) This case is displayed in Fig.~\ref{fig:iv-LEFT} and the integral
over the potential term (in the shaded area) reads
\begin{eqnarray}
&&\int_{t_0-h}^{t_0+h}\int_{x_0-h}^{x_p(t)}\,g_L(t,x)\,dt\,dx
=\int_{t_0-h}^{t_1}\,dt\int_{v_0-t-h}^{t+h-u_0}\,g_L(t,x)\,dx \nonumber\\
&&+\int_{t_1}^{t_0}\,dt\int_{v_0-t-h}^{x_p(t)}\,g_L(t,x)\,dx 
+\int_{t_0}^{t_2}\,dt\int_{t-h-u_0}^{x_p(t)}\,g_L(t,x)\,dx\nonumber\\
&&+\int_{t_2}^{t_0+h}\,dt\int_{t-h-u_0}^{v_0-t+h}\,g_L(t,x)\,dx.
\end{eqnarray}

\subsubsection{Right side integral}:

i) This case is displayed in Fig.~\ref{fig:i-LEFT} and the integral
over the potential term (in the non-shaded area) reads
\begin{eqnarray}
&&\int_{t_0-h}^{t_0+h}\int^{x_0+h}_{x_p(t)}\,g_R(t,x)\,dt\,dx
=\int_{t_0-h}^{t_1}\,dt\int_{v_0-t-h}^{t+h-u_0}\,g_R(t,x)\,dx \nonumber\\
&&+\int_{t_1}^{t_0}\,dt\int^{t+h-u_0}_{x_p(t)}\,g_R(t,x)\,dx 
+\int_{t_0}^{t_2}\,dt\int^{v_0-t+h}_{x_p(t)}\,g_R(t,x)\,dx\nonumber\\
&&+\int_{t_2}^{t_0+h}\,dt\int_{t-h-u_0}^{v_0-t+h}\,g_R(t,x)\,dx.
\end{eqnarray}

ii) This case is displayed in Fig.~\ref{fig:ii-LEFT} and the integral
over the potential term (in the non-shaded area) reads
\begin{eqnarray}
&&\int_{t_1}^{t_0+h}\int^{x_0+h}_{x_p(t)}\,g_R(t,x)\,dt\,dx
=\int_{t_1}^{t_0}\,dt\int^{t+h-u_0}_{x_p(t)}\,g_R(t,x)\,dx \nonumber\\
&&+\int_{t_0}^{t_2}\,dt\int^{v_0-t+h}_{x_p(t)}\,g_R(t,x)\,dx
+\int_{t_2}^{t_0+h}\,dt\int_{t-h-u_0}^{v_0-t+h}\,g_R(t,x)\,dx.
\end{eqnarray}

iii) This case is displayed in Fig.~\ref{fig:iii-LEFT} and the integral
over the potential term (in the non-shaded area) reads
\begin{eqnarray}
&&\int_{t_0-h}^{t_1}\int^{x_0+h}_{x_p(t)}\,g_R(t,x)\,dt\,dx
=\int_{t_0-h}^{t_1}\,dt\int_{v_0-t-h}^{t+h-u_0}\,g_R(t,x)\,dx \nonumber\\
&&+\int_{t_1}^{t_0}\,dt\int^{t+h-u_0}_{x_p(t)}\,g_R(t,x)\,dx +
\int_{t_0}^{t_2}\,dt\int^{v_0-t+h}_{x_p(t)}\,g_R(t,x)\,dx.
\end{eqnarray}

iv) This case is displayed in Fig.~\ref{fig:iv-LEFT} and the integral
over the potential term (in the non-shaded area) reads
\begin{eqnarray}
&&\int_{t_1}^{t_2}\int^{x_0+h}_{x_p(t)}\,g_R(t,x)\,dt\,dx\nonumber\\
&&=\int_{t_1}^{t_0}\,dt\int^{t+h-u_0}_{x_p(t)}\,g_R(t,x)\,dx +
\int_{t_0}^{t_2}\,dt\int^{v_0-t+h}_{x_p(t)}\,g_R(t,x)\,dx.
\end{eqnarray}

For all these cases we can perform explicitly the first of the two
integrals, then we need the explicit form of the trajectory to
proceed further.  We leave this for the direct numerical
implementation since explicit expressions are straightforward but
cumbersome.


\section{Implementation}

Here we will present the explicit implementation of the fourth order
accurate algorithm in vacuum, the initial data and boundary conditions
are also reviewed.

\subsection{Initial data}

While in the continuous Cauchy problem initial data of the second
order partial differential equation (\ref{Weq}) are specified by
$\psi(t=0,x)$ and $\partial_t\psi(t=0,x)$, a numerical code needs data
in previous time slices; in our case two. Assuming the field evolution
can be expanded in a Taylor series in time around $t=0$
\begin{equation}
\psi(t,x)=\psi(0,x)+\partial_t\psi(0,x)\,\left(\frac{t}{1!}\right)+
\partial_t^2\psi(0,x)\,\left(\frac{t^2}{2!}\right)+
\partial_t^3\psi(0,x)\,\left(\frac{t^3}{3!}\right)+
\partial_t^4\psi(0,x)\,\left(\frac{t^4}{4!}\right)+...,
\end{equation}
where second and higher order time derivatives can be computed using the
wave equation (\ref{Weq}), i.e.,
\begin{equation}
\partial_t^2\psi(0,x)=\partial_x^2\psi(0,x)-V_\ell(x)\,\psi(0,x)+S_{\ell\,m}(0,x),
\end{equation}
we can then obtain $\psi(h,x)$ to fourth order accurate in the
integration step $h$ thus providing the two time levels to start the
evolution algorithm.

Note that as pointed out in Ref.~\cite{Lousto97b}, this expansion
requires the particle not to cross the line joining grid points at
$t=0$ and $t=2h$. One can always choose carefully the location of the
(staggered-characteristic) grid points such that this never happens
for particles traveling at speeds less than that of the light.

\subsection{Boundary conditions}

The boundaries of the radial coordinate $x\,\dot=\,r^*$, that span the 
space outside the black hole from the horizon to spacial infinity,
are taken at a finite distance from the hole, $x_B$ and from the
event horizon $x_A$. At those boundaries we impose radiative
conditions as purely ingoing near the horizon and purely outgoing
at the outer boundary. These are clearly approximately true conditions
which improve as we push further the boundaries.

At the outer boundary we assume then the radiative fall off
\begin{equation}
\psi(u,x)\approx F_a(u)+\frac{F_b(u)}{x}+\frac{F_c(u)}{x^2}+...
\end{equation}

Evaluation of the above equation in three successive points along the constant $u$
direction give us the field at the boundary as a function of the two previous levels
\begin{eqnarray}
&&\psi(t+h,x_B)=\psi(t,x_B-h)+\\
&&\left[\psi(t,x_B-h)-\psi(t-h,x_B-2h)\right]
\left(1-\frac{2h}{x_B}\right)+{\cal O}\left(\frac{1}{x_B^2}\right)+{\cal O}\left(h^2\right).\nonumber
\end{eqnarray}

For the inner boundary, near the event horizon $x_A$ we have instead
\begin{equation}
\psi(v,x)\approx G_a(v)+\frac{G_b(v)}{x}+\frac{G_c(v)}{x^{2}}+...,
\end{equation}
which leads to
\begin{eqnarray}
&&\psi(t+h,x_A)=\psi(t,x_A+h)+\\
&&\left[\psi(t,x_A+h)-\psi(t-h,x_A+2h)\right]
\left(1+\frac{2h}{x_A}\right)+{\cal O}\left(\frac{1}{x_A^2}\right)+{\cal O}\left(h^2\right).\nonumber
\end{eqnarray}

As we see, the expansions near the boundaries not only depend on the
finite difference order, but also on the radiative condition that
only applies approximately. This is apparent in the power dependence
on the location of the numerical boundary and implies that pushing
the boundaries further away will reduce the amplitude of the
(spurious) reflected waves, but never completely eliminate its
effects. In order to do that, one should use the exact treatment of
the boundaries given in Ref.~\cite{Lau:2004as,Lau:2004jn}.

\subsection{Evolution in Vacuum}

Using Eqs.~(\ref{intg})-(\ref{cross}) we can make explicit how to update
the field $\psi$ using information on the two previous slices
\begin{eqnarray}
\psi(&&t+h,x)=-\psi(t-h,x)+\left[1+\frac14\left(\frac{h}{3}\right)^2\,V_\ell(x)\right]^{-1}
\Biggr\{\psi(t,x-h)+\psi(t,x+h)
\nonumber\\
&&-\frac14\left(\frac{h}{3}\right)^2\Biggr[V_\ell(x-h)\,\psi(t,x-h)+V_\ell(x+h)\,\psi(t,x+h)
+16\,V_\ell(x)\,\psi(t,x)
\nonumber\\
&&
+4\,\Bigg[V_\ell(x-h/2)\left(1-\frac12\left(\frac{h}{2}\right)^2\,V_\ell(x-h/2)\right)
\left(\psi(t,x-h)+\psi(t,x)\right)\nonumber\\
&&
+V_\ell(x+h/2)\left(1-\frac12\left(\frac{h}{2}\right)^2\,V_\ell(x+h/2)\right)
\left(\psi(t,x+h)+\psi(t,x)\right)
\Bigg]\Biggr]\Biggr\},
\end{eqnarray}
where $(t,x)$ are the coordinates of the center of each cell (not a grid point), and
$g_0=V_\ell(x)\,\psi(t,x)$ is given explicitly in Eq.~(\ref{g0c})-(\ref{g0r}).

To illustrate the numerical realization of the fourth order convergent
algorithm we implemented the above formulae in a numerical code to
compute waveforms as seen by an observer far away from the black holes
(in this example at $r^*=430M$.)  The initial distortion of the
Schwarzschild black hole is produced by a Gaussian, time-symmetric
perturbation
\begin{eqnarray}
\psi(t=0,r^*)&=&A\exp\left[-(r^*-r_c)^2/\sigma^2\right],\\
\partial_t\psi(t=0,r^*)&=&0,
\end{eqnarray}
where $A,r_c$ and $\sigma$ are parameters that we have taken here as $1.0,3.0,1.0$
respectively.

\begin{figure}
\begin{center}
\includegraphics[width=3.2in]{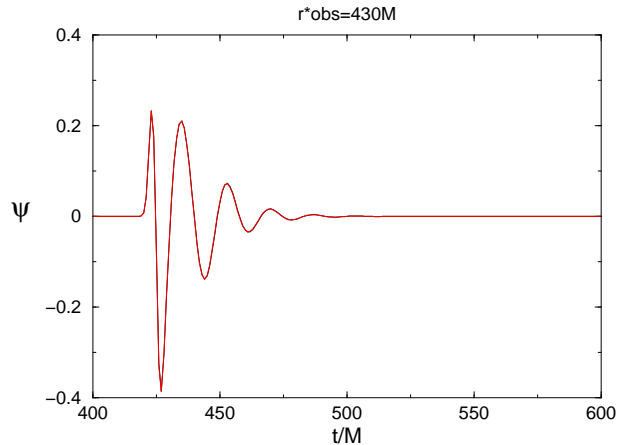}
\end{center}
\caption{The Waveform of a Gaussian, time-symmetric initial pulse as seen
by an observer located at $r^*=430M$. In this scale, second and fourth
order evolution waveforms lie on top of each other.}
\label{fig:wvfrm}
\end{figure}

\begin{figure}
\begin{center}
\includegraphics[width=3.0in]{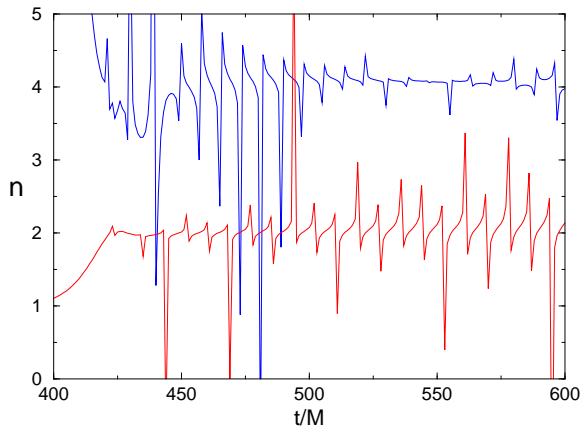}
\end{center}
\caption{The converging power of the fourth versus the second order algorithms.}
\label{fig:nWF}
\end{figure}

To check the convergence rate we computed waveforms at three different
resolutions $h,\ 2h$, and $4h$. In the example shown in Figs.~\ref{fig:wvfrm}
and \ref{fig:nWF}, we have taken $h=M/8$. We computed the convergence
rate simply as
\begin{equation}
n=\log\left|\frac{\psi(4h)-\psi(2h)}{\psi(2h)-\psi(h)}\right|/\log(2)
+\log\left|\epsilon^{(n)}(\xi)\right|/\log(2),
\end{equation}
where $\epsilon^{(n)}(\xi)$ represents the unknown error function of order $\approx1$.

The plots show the desired fourth order convergence rate on
average. For comparison we also implemented the second order algorithm
(\ref{vacumm2nd}) and displayed both rates together.


\section{Discussion}

Zerilli \cite{Zerilli70a} has written down explicitly all ten field
equations of General Relativity decomposed in tensor harmonics in the
Regge-Wheeler gauge \cite{Regge57}.  Then derived wave equations for
both degrees of freedom of the gravitational field, i.e. even and odd
parity perturbations, described in terms of (gauge invariant)
waveforms build up of metric perturbations and its first
derivatives. We described here how to integrate numerically in the
time domain these wave equation, when the perturbations are due to an
orbiting point-like source, represented by a Dirac's delta. In order
to complete the program we have to be able to reconstruct the metric
perturbations. This was done in Refs.~\cite{Lousto99b,Lousto:2005xu},
and applied to the radiation reaction problem in 
Refs.~\cite{Lousto99b,Lousto00a,BL02a}.

While we have a good understanding of perturbations due to an orbiting
particle around nonrotating black holes, we still need to make the
equivalent progress when dealing with perturbations of Kerr,
i.e. rotating black holes. The Teukolsky equation \cite{Teukolsky73}
describes those (curvature) perturbations compactly as a wave equation
for the Weyl scalars. It is straightforward to compute
\cite{Campanelli99} the energy and momentum radiated to
infinity by the system from the scalar $\Psi_4$. Yet, we need to
reconstruct the metric perturbations to compute, for instance the
force the particle exerts on itself and that accounts for the decay of
orbits due to gravitational radiation (See Refs.
\cite{2004LRR.....7....6P,Poisson:2004gg} for a review).
Recently this problem has been
tackled in the nonrotating limit
\cite{Lousto:2002em,Lousto:2005xu}. For the general Kerr background
the problem remains open [see Ref.~\cite{Ori:2002uv}.]

In any case, the starting point is to numerically solve, the Teukolsky
equation with an orbiting particle as a source. This plays very much
the role of the Zerilli and Regge-Wheeler equations that we just have
been dealing with. A first step has been recently
taken by generalizing the second order \cite{Krivan97a} to fourth
order
\cite{Pazos-Avalos:2004rp} accurate formalism, yet in vacuum. The
problem of generalizing the algorithms presented here for the orbiting
particle to the Teukolsky equation remains as another open question in
perturbation theory.


\ack
C.O.L. gratefully acknowledges the support of the NASA Center for
Gravitational Wave Astronomy at The University of Texas at Brownsville
(NAG5-13396), and for financial support from NSF grants PHY-0140326
and PHY-0354867.

\appendix

\section{Waveforms, Source terms and initial data}\label{apendiceA}

\subsection{Even parity source terms}

To obtain the Zerilli equation with sources for our waveform
\begin{equation}
\left[\partial_{r^*}^2-\partial_t^2-V_\ell^{even}(r)\right]
\psi_{even}^{\ell m}(r,t)=S_{even}^{\ell m}(r,t),
\label{zerillieq}
\end{equation}

we make the following combination of Einstein equations in the 
Regge-Wheeler gauge
[See Eq. (2.13) of Ref.~\cite{Lousto97a}]
\begin{eqnarray}
&&\frac{2(1-2M/r)}{r(\lambda+1)(\lambda r+3 M)}
\left[r^2 (1-2M/r)\partial_r R_{\theta\theta}\right.\nonumber\\
&&\left.
-(\lambda r+M) R_{\theta\theta}-\frac{\lambda r^4}{(\lambda r+3M)}
G_{tt}+r^3R_{tt}\right],
\end{eqnarray}

and find the source term is given by

\begin{eqnarray}
S_{even}^{\ell m}&=&8\pi\,{\frac { \,\left (r-2\,M\right )^{2}\left
(\lambda\,r-r+M\right )A_{\ell m}}{\left (\lambda+1\right )r\left
(\lambda\,r+3\,M\right )}}+16\pi\,{\frac {\,\left
(r-2\,M\right ) ^{2}B_{\ell m}}{\left (\lambda\,r+3\,M
\right )\left (\lambda+1\right )^{1/2}}}\nonumber\\
&&-8\pi\,{\frac {r\left({\lambda}^{2}{r}^{2}+8\,r\lambda\,M-{r}^{2}
\lambda-9\,rM+27\,{M}^{2}\right ) \,{A^{(0)}_{\ell m}}}{\left (\lambda+1\right )\left(\lambda\,r+3\,M\right )^{2}}}\nonumber\\
&&+8\pi\,{\frac {\,\left (r-2\,
M\right )^{2}\sqrt {2}\,G_{\ell m}}{\left (\lambda+1\right )\left (\lambda\,r+
3\,M\right )}}
-8\pi\,{\frac {\,\left (r-2\,M\right )\sqrt {2}\,F_{\ell m}}
{\sqrt {\lambda\,\left (\lambda+1\right )}}}\nonumber\\
&&-8\pi\,{\frac {\left (r-2\,M\right )^{3} \,\partial_r A_{\ell m}}{
\left (\lambda+1\right )\left (\lambda\,r+3\,M\right )}}
+8\pi\,{\frac{\left (r-2\,M\right ){r}^{2}
\,{\partial_r {A^{(0)}_{\ell m}}}}{\left (
\lambda+1\right )\left (\lambda\,r+3\,M\right )}}.
\end{eqnarray}

Note that we had to reverse the sign of the (even) $t\,\theta$ component
of the Einstein equations as given by Zerilli \cite{Zerilli70a}. Thus
the source term proportional to $B^{(0)}_{\ell m}$ changes sign.

The source term of the Zerilli equation (\ref{zerillieq}) has
now the following form

\begin{equation}
S_{even}^{\ell m}=F(t)\,\frac{\partial}{\partial_r}\delta[r-R(t)]+G(t)\,\delta[r-R(t)],
\end{equation}
where
\begin{eqnarray}
&&F_{even}(t)=-8\pi\,{\frac {{m_0}\,{U^0}\left (R-2\,M\right )\left (\left (
{\frac {d}{dt}}R\right )^{2}{R}^{2}-\left (R-2\,M\right )^{2}\right ) 
}{\left (\lambda+1\right )\left (\lambda\,R+3\,M
\right ){R}^{2}}}\bar{Y}_{\ell m},\nonumber\\
\nonumber\\
\nonumber\\
&&G_{even}(t)= 16\,{\frac {\left ({\frac {d}{dt}}R\right )\pi \,m_0\,U^0\left
(R-2\,M\right ){\frac {d}{dt}}\bar{Y}_{\ell m}}{\left (\lambda
R+3\,M\right )\left (
\lambda+1\right )}}\nonumber\\
&&-8\,{\frac {\left (R-2\,M\right )U^0{m_0}\,\pi \,\bar{X}_{\ell
m}\left ({\frac {d}{dt}}\Theta\right ){\frac
{d}{dt}}\Phi}{\lambda\,\left (\lambda+1\right )}}\nonumber\\
&&-4\,{\frac {\left (R-2\,M\right )U^0m_0\,\pi \,\left (\left ({\frac
{d}{dt}}
\Theta\right )^{2}-\left (\sin(\theta)\right )^{2}\left ({\frac {d}
{dt}}\Phi\right )^{2}\right )\bar{W}^{\ell m}}{\lambda\,\left (\lambda+1
\right )}}\nonumber\\
&&+8\,{\frac {\bar{Y}_{\ell m}\pi \,m_0\,U^0\left (R-2\,M\right
)^{2}\left (\left ({\frac {d}{dt}}\Theta\right )^{2}+\left
(\sin(\theta)\right )^{2}\left ({\frac {d}{dt}}\Phi\right
)^{2}\right )}{\left (\lambda R+3\,M\right )\left (\lambda+1\right
)}}\nonumber\\ &&+8\,{\frac {\bar{Y}_{\ell m}\pi \,m_0\,U^0\left
({R}^{2}\lambda+6\,RM+6\,\lambda
R\,M+3\,{M}^{2}+{R}^{2}{\lambda}^{2}\right )\left (\left ({\frac
{d}{dt}}R\right )\right )^{2}}{\left (\lambda R+3\,M\right )^{2}\left
(\lambda+1\right )R}}\nonumber\\ &&-8\,{\frac {\bar{Y}_{\ell m}\pi
\,m_0\,U^0\left ( R-2\,M\right )^{2}\left
({R}^{2}\lambda+{R}^{2}{\lambda}^{2}+15\,{M}^{2}+6\,\lambda R\,M\right
)}{{R}^{3}\left (\lambda+1\right )\left (\lambda R+3\,M\right )^{2}}},
\end{eqnarray}
and where $R,\Theta,\Phi$ give the trajectory of the orbiting particle 
in spherical coordinates and $W^{\ell m}$ is defined in Eq.~(\ref{W}).


\subsection{Even Parity Initial data}

In Refs.~\cite{Lousto97b,Lousto98a} we have introduced a conformally
flat index as a combination of metric perturbations
\begin{equation}\label{confindex}
I_{\rm conf}^{even}\equiv
H_2^{\ell m}-K^{\ell m}+\frac{2}{r}\left(1-\frac{3M}{r}
\right)\ \left(
h_1^{\ell m}-\frac{r^2}{2}\partial_rG^{\ell m}
\right)
-2\left(1-\frac{2M}{r}
\right)\partial_r\left(
h_1^{\ell m}-\frac{r^2}{2}\partial_rG^{\ell m}
\right)
\ .
\end{equation} 
which is gauge invariant, and clearly vanishes for a 3-geometry that
is in conformally flat form, with $h_1^{\ell m}=0, G^{\ell m}=0$ and $H_2^{\ell m}=K^{\ell m}$. The
computation of this gauge invariant quantity from $\psi_{even}^{\ell m}(r)$ is most
easily described in the Regge-Wheeler gauge, where $I_{\rm conf}^{even}$ reduces to
$H_2^{\ell m}-K^{\ell m}$.

Now making use of this $I_{\rm conf}^{even}$ in the Regge-Wheeler gauge and
the expressions \cite{Lousto:2005xu} for $H_{2}^{\ell m}$ and 
$K^{\ell m}$ in terms of the waveform $\psi_{even}^{\ell m}$ we can write
\begin{eqnarray}\label{energia}
&&I_{\rm conf}^{even}
=(r-2M)\partial^2_r\psi_{even}^{\ell m}+\frac{[(\lambda-3)r+9M]M}
{(\lambda r+3M)r}\partial_r\psi_{even}^{\ell m}\nonumber\\
&&-\frac{[27M^3+24\lambda M^2r+3\lambda(3\lambda+1)Mr^2+2\lambda^2
(\lambda+1)r^3]}{(\lambda r+3M)^2r^2}\psi_{even}^{\ell m}\\
&&+\frac{\kappa\ U^0(1-2M/r)
[\lambda(\lambda+1)r^2-3M^2]}{(\lambda+1)(\lambda r+3M)^2}\delta[r-R]
-\frac{\kappa\ U^0(r-2M)^2}{(\lambda+1)(\lambda r+3M)}\partial_r\delta[r-R],\nonumber
\end{eqnarray}
where $\kappa=8\pi m_0 Y_{\ell m}[\Theta(t),\Phi(t)]$.  This
equation is completely equivalent to the Hamiltonian constraint
\cite{Lousto:2005xu}. It is only written in terms of the two variables
$I_{\rm conf}^{even}$ and $\psi_{even}^{\ell m}$ instead of $H_{2}^{\ell m}$
and $K^{\ell m}$.  The conformally flat initial data that we studied
in Refs.~\cite{Lousto98a} corresponds to choose $I_{\rm conf}^{even}=0$ and
solve (with specified boundary conditions) for the resulting
differential equation for $\psi_{even}^{\ell m}$.

Interestingly enough, the time derivative (represented by an overdot)
of this equation
\begin{eqnarray}\label{momentum}
\dot I_{\rm conf}^{even}&=&(r-2M)\partial_r^2\dot\psi_{even}^{\ell m}+
\frac{[(\lambda-3)r+9M]M}{(\lambda r+3M)r}\partial_r\dot\psi_{even}^{\ell m}\nonumber\\
&&-\frac{[27M^3+24\lambda M^2r+3\lambda(3\lambda+1)Mr^2+2\lambda^2
(\lambda+1)r^3]}{(\lambda r+3M)^2r^2}\dot\psi_{even}^{\ell m}\nonumber\\
&&+\frac{2\kappa\ U^0\dot R
[9M^2+2M(\lambda-3)r-\lambda(\lambda+3)r^2]} {(\lambda+1)r^2(\lambda
r+3M)^2}\delta[r-R]\nonumber\\ &&+\frac{8\pi m_0 (d\bar{Y}_{\ell
m}/dt) U^0 (r-2M)(\lambda(\lambda+1)r^2-3M^2)} {(\lambda+1)r(\lambda
r+3M)^2}\delta[r-R]\nonumber\\
&&+\frac{\kappa\,U^0\dot{r}_p(r-2M)[9M^2+2M\lambda
r-\lambda(\lambda+1)r^2]} {(\lambda+1)r(\lambda
r+3M)^2}\partial_r\delta[r-R]\nonumber\\ &&-\frac{8\pi m_0 (d\bar{Y}_{\ell m}/dt)
U^0 (r-2M)^2}{(\lambda+1) (\lambda r+3M)}\partial_r\delta[r-R]\nonumber\\
&&+\frac{\kappa\ U^0\dot R(r-2M)^2} {(\lambda+1)(\lambda
r+3M)}\partial_r^2\delta[r-R]\ .
\end{eqnarray}
is equivalent to the equations for the momentum constraint, i.e.
when we replace the metric perturbations $K^{\ell m},H_2^{\ell m},H_1^{\ell m}$ into
{\it either} the $(tr)$ or $(t\varphi)$ components of the 
General Relativity constraints \cite{Lousto:2005xu}, both yield Eq.~(\ref{momentum}) above.

By specifying conditions on $I_{\rm conf}$ and $\dot I_{\rm conf}$ on
an initial hypersurface $\Sigma$ one can determine $\psi^{\ell
m}_{t_\Sigma}$ and $\dot \psi^{\ell m}_{t_\Sigma}$ which is the
initial data we need to evolve Zerilli's equation. In
Ref.~\cite{Lousto98a} we have taken
\begin{equation}
{I}_{\rm conf}|_{t_\Sigma}=0,~~\dot{I}_{\rm conf}|_{t_\Sigma}=0\ .
\end{equation}
to determine the ``convective initial data
\begin{equation}\label{psiprop}
\psi^{\ell m}_{even}|_{t_\Sigma}=\psi^{\ell m}_{even}(x^k;x^k_p[t_\Sigma]) 
\end{equation}\begin{equation}\label{psidotprop}
\dot\psi^{\ell m}_{even}|_{t_\Sigma}=\left.\frac{\partial\psi^{\ell m}_{even}(x^k;x^k_p[t])}
{\partial t}\right|_{t_\Sigma}
=\left[\frac{dx^j_p[t]}{dt}\ 
\frac{\partial\psi^{\ell m}_{even}(x^k;x^k_p[t])}{\partial x^j_p}\right]_{t_\Sigma}\ .
\end{equation}

For the case of circular orbits this data takes the form
\begin{eqnarray}\label{psi0}
&&\psi^{\ell m}_{even}|_{t_\Sigma}=\frac{2m(R)}{\lambda+1}\frac{
\sqrt{4\pi/(2\ell+1)}}{\lambda r+3M}r\sqrt{r/\bar{r}}\\
\nonumber\\
&&\times
\begin{cases}
\left(\lambda+1+M/r+\sqrt{1-2M/r}\left(\ell+\sqrt{\bar{r}/r}\right)
\right)(\bar{R}/\bar{r})^\ell\quad &;\bar{r}>\bar{R},\\
\nonumber\\
\left(\lambda+1+M/r+\sqrt{1-2M/r}\left(\sqrt{\bar{r}/r}-\ell-1\right)
\right)(\bar{r}/\bar{R})^{\ell+1}\quad &;\bar{r}<\bar{R},
\end{cases}
\end{eqnarray}
where $\bar{r}=1/4(\sqrt{r}+\sqrt{r-2M})^2$ is the isotropic coordinate.

\begin{eqnarray}
\dot\psi^{\ell m}_{even}|_{t_\Sigma}=\left.-\left(\frac{d\Phi}{dt}\right)
\frac{\partial\psi^{\ell m}_{even}}{\partial\Phi}\right|_{t_\Sigma}\ ,
=\left.-im\Omega\,\psi^{\ell m}_{even}\right|_{t_\Sigma}\ ,
\end{eqnarray}

\subsection{Odd parity source term}

The Hilbert-Einstein equations in the Regge-Wheeler {\it gauge} for
the odd parity sector (See Zerilli's\cite{Zerilli70a} equations
(C6a)-(C6c)) [Note the corrections to the source terms]
\begin{eqnarray}
&&\frac{\partial^2h_0^{\ell m}}{\partial r^2}-\frac{\partial^2h_1^{\ell m}}
{\partial r\partial t} -\frac2r\frac{\partial h_1^{\ell m}}{\partial t}+
\left[\frac{4M}{r^2}-\frac{2(\lambda+1)}{r}\right]
\frac{h_0^{\ell m}}{r-2M}\nonumber\\
&=&-\frac{8\pi\,rQ^{(0)}_{\ell m}}{(1-2M/r)\sqrt{(\lambda+1)}}\label{tphiodd},\\
\nonumber\\
&&\frac{\partial^2h_1^{\ell m}}{\partial t^2}-\frac{\partial^2h_0^{\ell m}}{\partial
r\partial t} +\frac2r\frac{\partial h_0^{\ell m}}{\partial t}+2\lambda
(r-2M)\frac{h_1^{\ell m}}{r^3}\nonumber\\
&=&\frac{8\pi\,i(r-2M)Q_{\ell m}}{\sqrt{(\lambda+1)}}\label{rphiodd},\\
\nonumber\\
&&(1-2M/r)\frac{\partial h_1^{\ell m}}{\partial r}-\frac{1}{(1-2M/r)}
\frac{\partial h_0^{\ell m}}{\partial t}
+\frac{2M}{r^2}h_1\nonumber\\
&=&-\frac{4\pi\,ir^2D_{\ell m}}{\sqrt{\lambda(\lambda+1)}},
\label{thetaphiodd}
\end{eqnarray}
where $Q^{(0)}_{\ell m}, Q_{\ell m}$ and $D_{\ell m}$ give the multipole
decomposition of the energy-momentum tensor.

One can use the above equations to write the metric perturbation in the
Regge--Wheeler {\it gauge}
\begin{eqnarray}
h_0^{\ell m}(r,t)&=&\frac12(1-2M/r)\partial_r\left(r\psi^{\ell
m}\right)\label{h0odd}+\frac{4\pi r^3Q^{(0)}_{\ell
m}}{\lambda\sqrt{(\lambda+1)}},\\
\nonumber\\
h_1^{\ell m}(r,t)&=&\frac12\frac{r}{(1-2M/r)}\partial_t\psi^{\ell
m}\label{h1odd}+\frac{4\pi i r^3 Q_{\ell
m}}{\lambda\sqrt{(\lambda+1)}},
\end{eqnarray}

Taking the time derivative of Eq.\ (\ref{rphiodd}) and the radial
derivative of Eq.\ (\ref{tphiodd}) allow us to reconstruct the
Regge-Wheeler equation for odd parity perturbations
\begin{equation}
\left[\partial_{r^*}^2-\partial_t^2-V_\ell^{odd}(r)\right]
\psi^{\ell m}(r,t)=S_{odd}^{\ell m}(r,t),
\label{rweq}
\end{equation}
where the source is given by
\begin{eqnarray}
S_{odd}^{\ell m}=-{\frac {8\pi \,\left (r-2\,M\right
)}{\lambda\sqrt {\left (\lambda+1\right )}}}
\left [{\frac {
\partial }{\partial r}}\left(r{Q^{(0)}_{\ell m}}(r,t)\right)-ir{\frac {\partial }{
\partial t}}Q(r,t)_{\ell m}
\right].
\end{eqnarray}

For the Zerilli's variable $\psi^{\ell m}_{Zer,odd}$ the source term
looks like instead
\begin{eqnarray}
S_{Zer,odd}^{\ell m}&=&\frac{8\pi\,i(1-2M/r)}{\sqrt{\lambda+1}}
\left[-(1-2M/r)\,Q_{\ell m}+\frac{r}{\sqrt{2\lambda}}
\partial_r\left((1-2M/r)D_{\ell m}\right)\right].\nonumber\\
\end{eqnarray}

For a source term represented by a particle
\begin{eqnarray}
Q^{(0)}_{\ell m}&=&\frac{m_0\,U^0(t)(1-2M/r)\,{\rm ang}(t)\,\delta[r-R(t)]}
{r\sqrt{\lambda+1}},\\
\nonumber\\
Q_{\ell m}&=&\frac{i\,m_0\,U^0(t)(\frac{d}{dt}R)\,{\rm ang}(t)\,\delta[r-R(t)]}
{(r-2M)\sqrt{\lambda+1}},\\
\nonumber\\
D_{\ell m}&=&-\frac{i\,m_0\,U^0(t)\,{\rm ang2}(t)\,\delta[r-R(t)]}
{\sqrt{2\lambda(\lambda+1)}},
\end{eqnarray}
where
\begin{equation}
{\rm ang}(t)=
\frac{1}{\sin\Theta}\left(\frac{d\Theta}{dt}\right)
\partial_\varphi\bar{Y}^{\ell m}
(\Theta,\Phi)-\sin\Theta\left(\frac{d\Phi}{dt}\right)
\partial_\theta\bar{Y}^{\ell m}(\Theta,\Phi),
\end{equation}
\begin{eqnarray}
{\rm ang2}(t)&=&\frac12\left[\left(\frac{d\Theta}{dt}\right)^2
-\sin^2\Theta\left(\frac{d\Phi}{dt}\right)^2\right]\nonumber\\
&\times&\left[\frac{1}{\sin\Theta}\bar{X}^{\ell m}\right]
-\sin\Theta\frac{d\Phi}{dt}\frac{d\Theta}{dt}\bar{W}^{\ell m},
\end{eqnarray}
and $R,\Theta,\Phi$ define the trajectory of the orbiting particle 
in spherical coordinates. We also used Zerilli's notation
\begin{eqnarray}
{X}^{\ell m}&=&2\partial_\varphi\Big(\partial_\theta-\cot\theta\Big)
Y^{\ell m},\label{X}\\
\nonumber\\
{W}^{\ell m}&=&\left(\partial^2_\theta-\cot\theta\,\partial_\theta-
\frac{1}{\sin^2\theta}\partial^2_\varphi\right)Y^{\ell m}.\label{W}
\end{eqnarray}

The source term of the Regge-Wheeler equation (\ref{rweq}) has
now the following form

\begin{equation}
S_{odd}^{\ell m}
=F(t)\,\frac{\partial}{\partial_r}\delta[r-R(t)]+G(t)\,\delta[r-R(t)],
\end{equation}
where
\begin{eqnarray}
F(t)&=&-{\frac {8\pi \,{m_0}\,{U^0}(t)
\left[(R-2M)^2-R^2(\partial_tR)^2\right]}
{\lambda(\lambda+1)\,R}}{\rm ang}(t),\\
\nonumber\\
\nonumber\\
G(t)=&-&{\frac {8\pi{m_0}\,{U^0}(t)R\left ({\frac {d}{dt}}R\right
)\left ({\frac {d}{dt}}{\rm ang}(t)\right )}{\lambda(\lambda+1)}}
-{\frac {8\pi{m_0} \,{\rm ang}(t)}{\lambda(\lambda+1)R}}\nonumber\\
&\times&\Bigg\{
R^{2}\left ({\frac {d}{dt}}{ U^0}(t)\right ){
\frac {d}{dt}}R+R^{2}{ U^0}(t){\frac {d^{2}}{d{
t}^{2}}}R\nonumber\\
&+&2M\,{U^0}(t)-R\,{U^0}(t)+R\,{U^0}(t)\left ({
\frac {d}{dt}}R\right )^{2}
\Bigg\}.
\end{eqnarray}

\subsection{Odd Parity Initial data}

In analogy to the 'even' conformally flat index 
we define the following combination of metric perturbations
\begin{equation}\label{confindex-Odd}
I_{\rm conf}^{Odd}\equiv
h_1^{\ell m}+\frac{r^2}{2}\partial_r\left(\frac{h_2^{\ell m}}{r^2}\right)\ .
\end{equation} 
which is gauge invariant, and clearly vanishes for a 3-geometry that
is in conformally flat form. With $h_2^{\ell m}=0$ in the Regge-Wheeler gauge,
an initially conformally flat metric 
$\left.I_{\rm conf}^{Odd}\right|_{\Sigma}=0$ means
\begin{equation}\label{h1=0}
\left.h_1^{\ell m}\right|_{\Sigma}=0.
\end{equation} 

Now taking also $\left.\dot I_{\rm conf}^{Odd}\right|_{t_{\Sigma}}=0$ in the
Regge-Wheeler gauge leads to
\begin{equation}\label{doth1=0}
\left.\partial_t h_1^{\ell m}\right|_{t_\Sigma}=0,
\end{equation} 
which is the Odd parity version of the CF thin-sandwich data with
$\tilde{u}_{ij}=0$ (See Ref.~\cite{York99,Cook:2001wi}). While $\dot
I=0$ is the Wilson-Mathew ~\cite{Marronetti:1999ya} approach adapted
to black hole initial data. Other equivalent (at our perturbative
level) formulation was given
by~\cite{Gourgoulhon02,Grandclement02} proposing a helical
Killing vector symmetry.

From Eq.~(\ref{h0odd}) we see that (\ref{h1=0}) gives
\begin{equation}\label{dotpsi=0}
I_{\rm conf}^{Odd}=0 =>
\left.\partial_t\psi^{\ell m}_{Odd}\right|_{t_\Sigma}=0,
\end{equation} 
and from the definition of the waveform~(\ref{psiodd}) the condition
$\dot h_1^{\ell m}=0$ gives
\begin{equation}\label{psit=0}
\partial_t I_{\rm conf}^{Odd}=0 =>
\left.\psi^{\ell m}_{Odd}\right|_{t_\Sigma}=\frac{r^3}{\lambda}
\partial_r\left(\frac{h_0^{\ell m}}{r^2}\right) .
\end{equation} 
The momentum constraint (\ref{tphiodd}) then gives us the form of
$h_0|_{t_\Sigma}$

\begin{eqnarray}
&&\frac{\partial^2h_0}{\partial \xi^2}+
\left[\frac{2}{\xi^2}-\frac{\ell(\ell+1)}{\xi}\right]
\frac{h_0}{\xi-1}
=-\frac{8\pi\,m_0\,U^0(t_\Sigma)2M\,ang(t_\Sigma)}
{(\lambda+1)}\delta(\xi-\xi_p),\nonumber\\
\label{h0init}
\end{eqnarray}
where $\xi=r/(2M)$ and $\xi_p=R(t)/(2M)$.

The solutions to this equation can be written in terms of hypergeometric
functions
\begin{eqnarray}
h_0^{\ell m}|_{t_\Sigma}=
\begin{cases}
{\rm C_1(\xi_p)}y_1(\xi)~,~~\xi\leq \xi_p,\\
{\rm C_2(\xi_p)}y_2(\xi)~,~~\xi\geq \xi_p,\\
\end{cases}
\end{eqnarray}
[Here we have the freedom of adding $y_2$ for all $\xi$ as an homogeneous
solution representing additional 'radiation content'].
where
\begin{eqnarray}
y_1(1)&=&{\xi}^{2} \left( \xi-1 \right) ^{-1+\ell}\,_{1}F_{2}\left( [
1-\ell,2-\ell],[-2\,\ell],- \left( \xi-1 \right) ^{-1}
\right),\\
y_2(\xi)&=&{\xi}^{2} \left( \xi-1 \right) ^{-2-\ell}\,_{1}F_{2}
\left( [\ell+3,\ell+2],[2+2\,\ell],- \left( \xi-1 \right) ^{-1} \right).
\end{eqnarray}

In order to determine the value of the constant ${\rm C(\xi_p)}$ we
ensure the satisfaction of Eq.~(\ref{h0init}) at $r=R$ by equating
the jump the the first derivatives of $h_0$ with the coefficient of
the $\delta(\xi-\xi_p)$ in the source term of Eq.~(\ref{h0init}). The
result is
\begin{equation}
C_{1,2}(\xi_p)=\left.8\pi\,m_0\,U^0(R)\,{\rm ang}^{\ell m}
\left[\Theta(t_\Sigma),\Phi(t_\Sigma)\right]\frac{2M}
{(\lambda+1)}\frac{y_{1,2}}{W}\right|_{(\xi_p)}
\end{equation}
with $W=y_1'\,y_2-y_2'\,y_1=2\ell+1$.

For instance, for $\ell=2$ we get

\begin{eqnarray}
\left.\psi^{\ell=2,m}\right|_{t_\Sigma}=
\begin{cases}
{\rm C_1}\,\frac{\xi^3}{\lambda}~,~~\xi\leq \xi_p,\\
\\
{\rm C_2}\,\left[-\frac{20\xi^3}{\lambda}\ln\left|\frac{\xi}{\xi-1}\right|+
\frac{5(12\xi^3+6\xi^2+4\xi+3)}{3\lambda \xi}\right]~,~~\xi\geq \xi_p.
\end{cases}
\end{eqnarray}

For $\ell=3$ we get

\begin{eqnarray}
\left.\psi^{\ell=2,m}\right|_{t_\Sigma}=
\begin{cases}
{\rm C_1}\,\frac{\xi^3}{\lambda}(2\xi-5/3)~,~~\xi\leq \xi_p,\\
\\
{\rm C_2}\,\left[-\frac{210\xi^3(6\xi-5)}{\lambda}\ln\left|\frac{\xi}{\xi-1}\right|+
\frac{7(360\xi^4-120\xi^3-30\xi^2-10\xi-3)}{2\lambda \xi}\right]~,~~\xi\geq \xi_p.
\end{cases}
\end{eqnarray}
Successive multipoles can be computed in the same way when needed.


\section{Numerical construction of up to second order derivatives}\label{appendix:g}

In Sec.~\ref{Sec:particle} we have assumed that we can Taylor expand
the function $g(t,x)=V(x)\,\psi(t,x)$ around the left and right corners
of the cell (centered at $(t_0,x_0))$ the particle crosses
\begin{eqnarray}\label{Taylorbis}
g_{R,L}(t,x)&=&g[t_0,x_0\pm h]+\frac{\partial g}{\partial x}[t_0,x_0\pm h]\,(x-x_0\mp h)+\nonumber\\
&&\frac{\partial g}{\partial t}[t_0,x_0\pm h]\,(t-t_0)+
\frac{\partial^2 g}{\partial x^2}[t_0,x_0\pm h]
\,\frac{(x-x_0\mp h)^2}{2!}+\nonumber\\
&&\frac{\partial^2 g}{\partial t\partial x}[t_0,x_0\pm h]\,(t-t_0)\,(x-x_0\mp h)+\nonumber\\
&&\frac{\partial^2 g}{\partial t^2}[t_0,x_0\pm h]
\,\frac{(t-t_0)^2}{2!}+{\cal O}\left(h^3\right).
\end{eqnarray}
So, for numerical purposes, we need
to construct the derivatives appearing here in terms of the function
evaluated in nearby grid points. The technique we use is to evaluate
the Taylor expansion above (\ref{Taylorbis}) in six nearby point to
evaluate $g$ and all up to second derivatives of $g$ at the left/right
corner of the cell the particle crosses through.

\subsection{}
\begin{figure}
\begin{center}
\includegraphics[width=5.0in]{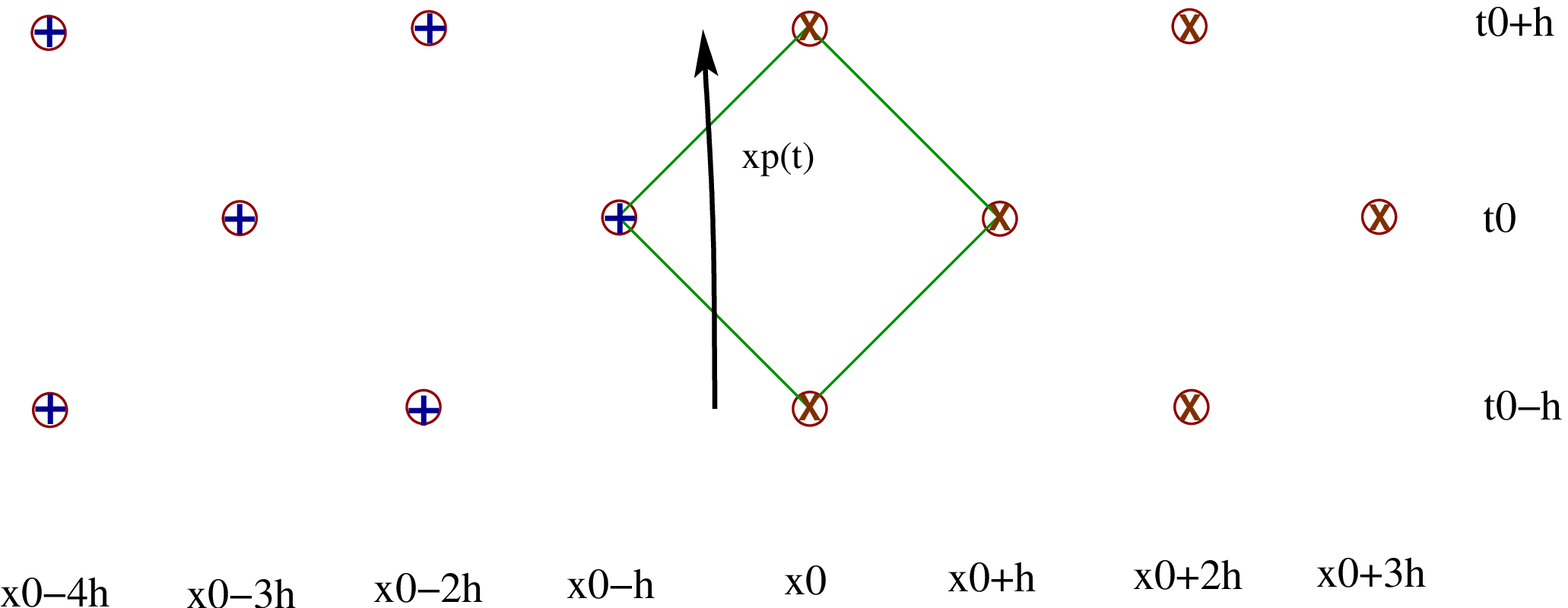}
\end{center}
\caption{Case i) the particle enters and leaves to the left. The
field $\psi$ is approximated by an expansion around the left corner
of the cell using the points labeled by $\oplus$ and around the
right corner by points labeled by $\otimes$.}
\label{fig:UNO-LEFT}
\end{figure}

Case i), the particle enters and leaves the cell on the left side. The
construction of the derivatives is made taking into account the points
labeled with the crossed circle on the left side, $\oplus$, and with
the Xed-circle on the right, $\otimes$ [See Fig.~\ref{fig:UNO-LEFT}.]  To
the required order, the derivatives in terms of grid points
evaluations read explicitly

For the integration on the left

\begin{eqnarray}\label{i-left-deriv}
\frac{\partial g_L}{\partial x}[t_0,x_0-h]=
&&\frac{4g(t_0,x_0-h)+g(t_0+h,x_0-4h)+g(t_0-h,x_0-4h)}{4h}\nonumber\\
&&-\frac{4g(t_0,x_0-3h)+g(t_0-h,x_0-2h)+g(t_0+h,x_0-2h)}{4h},\nonumber\\&&
\\
\frac{\partial g_L}{\partial t}[t_0,x_0-h]=
&&\frac{3g(t_0+h,x_0-2h)-3g(t_0-h,x_0-2h)}{4h}\nonumber\\
&&\frac{-g(t_0+h,x_0-4h)+g(t_0-h,x_0-4h)}{4h},
\\
\frac{\partial^2 g_L}{\partial x^2}[t_0,x_0-h]=
&&\frac{g(t_0+h,x_0-4h)+2g(t_0,x_0-h)+g(t_0-h,x_0-4h)}{4h^2}\nonumber\\
&&-\frac{g(t_0-h,x_0-2h)+g(t_0+h,x_0-2h)+2g(t_0,x_0-3h)}{4h^2},\nonumber\\&&
\\
\frac{\partial^2 g_L}{\partial t\partial x}[t_0,x_0-h]=
&&\frac{g(t_0-h,x_0-4h)-g(t_0-h,x_0-2h)}{4h^2}\nonumber\\
&&\frac{-g(t_0+h,x_0-4h)+g(t_0+h,x_0-2h)}{4h^2},
\\
\frac{\partial^2 g_L}{\partial t^2}[t_0,x_0-h]=
&&\frac{3g(t_0-h,x_0-2h)-2g(t_0,x_0-h)}{4h^2}\nonumber\\
&&+\frac{g(t_0+h,x_0-4h)+g(t_0+h,x_0-2h)}{4h^2}\nonumber\\
&&+\frac{g(t_0-h,x_0-4h)+3g(t_0+h,x_0-2h)-6g(t_0,x_0-3h)}{4h^2},\nonumber\\&&
\end{eqnarray}

and for the integration on the right

\begin{eqnarray}\label{i-right-deriv}
\frac{\partial g_R}{\partial x}[t_0,x_0+h]=
&&\frac{-g(t_0+h,x_0+2h)+g(t_0-h,x_0)}{4h}\nonumber\\
&&+\frac{g(t_0+h,x_0)-g(t_0-h,x_0+2h)}{4h},
\\
\frac{\partial g_R}{\partial t}[t_0,x_0+h]=
&&\frac{g(t_0+h,x_0)-g(t_0-h,x_0+2h)}{4h}\nonumber\\
&&\frac{-g(t_0-h,x_0)+g(t_0+h,x_0+2h)}{4h},
\\
\frac{\partial^2 g_R}{\partial x^2}[t_0,x_0+h]=
&&\frac{-g(t_0+h,x_0+2h)-2g(t_0,x_0+h)+g(t_0-h,x_0)}{4h^2}\nonumber\\
&&+\frac{g(t_0+h,x_0)-g(t_0-h,x_0+2h)+2g(t_0,x_0+3h)}{4h^2},
\nonumber\\
&&\\
\frac{\partial^2 g_R}{\partial t\partial x}[t_0,x_0+h]=
&&\frac{-g(t_0-h,x_0)+g(t_0+h,x_0)}{4h^2}\nonumber\\
&&\frac{-g(t_0+h,x_0+2h)+g(t_0-h,x_0+2h)}{4h^2},
\\
\frac{\partial^2 g_R}{\partial t^2}[t_0,x_0+h]=
&&\frac{3g(t_0-h,x_0+2h)-6g(t_0,x_0+h)-2g(t_0,x_0+3h)}{4h^2}\nonumber\\
&&+\frac{g(t_0-h,x_0)+g(t_0+h,x_0)+3g(t_0+h,x_0+2h)}{4h^2}.
\nonumber\\
\end{eqnarray}

\subsection{}
\begin{figure}
\begin{center}
\includegraphics[width=5.0in]{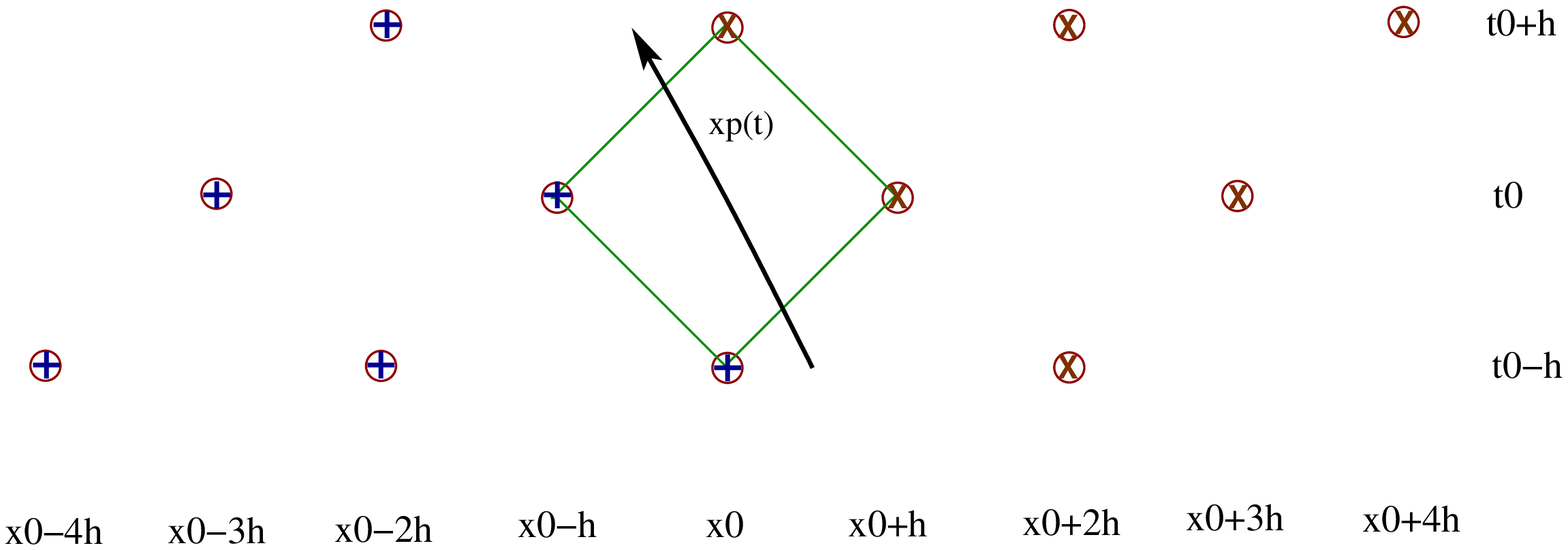}
\end{center}
\caption{Case ii) the particle enters the cell on the right and leaves
it on the left. The field $\psi$ is approximated by an expansion
around the left corner of the cell using the points labeled by
$\oplus$ and around the right corner by points labeled by $\otimes$.}
\label{fig:DOS-LEFT}
\end{figure}

Case ii), the particle enters the cell on the right side and leaves it
on the left side. The construction of the derivatives is made taking
into account the points labeled with the crossed circle on the left
side, $\oplus$, and with the Xed-circle on the right, $\otimes$ [See
Fig.~\ref{fig:DOS-LEFT}.]  Note that if we had chosen the alternative
point $g(t_0+h,x_0-4h)$ instead of $g(t_0-h,x_0-4h)$ to the left, the
set of equations do not produce a solution.  The same applies to the
right of the particle if we had taken $g(t_0-h,x_0+4h)$ instead of
$g(t_0+h,x_0+4h)$ as the sixth grid point.

To the required order, the
derivatives in terms of grid points evaluations read explicitly
for the integration on the left

\begin{eqnarray}\label{ii-left-deriv}
\frac{\partial g_L}{\partial x}[t_0,x_0-h]=
&&\frac{-2g(t_0,x_0-3h)+2g(t_0,x_0-h)-2g(t_0-h,x_0-2h)}{4h}\nonumber\\
&&+\frac{g(t_0-h,x_0)+g(t_0-h,x_0-4h)}{4h},
\\
\frac{\partial g_L}{\partial t}[t_0,x_0-h]=
&&\frac{-2g(t_0-h,x_0-2h)-g(t_0-h,x_0)+2g(t_0,x_0-h)}{4h}\nonumber\\
&&\frac{-2g(t_0,x_0-3h)+g(t_0-h,x_0-4h)+2g(t_0+h,x_0-2h)}{4h},
\\
\frac{\partial^2 g_L}{\partial x^2}[t_0,x_0-h]=
&&\frac{-2g(t_0-h,x_0-2h)+g(t_0-h,x_0)+g(t_0-h,x_0-4h)}{4h^2},
\\
\frac{\partial^2 g_L}{\partial t\partial x}[t_0,x_0-h]=
&&\frac{-g(t_0-h,x_0)+2g(t_0,x_0-h)-2g(t_0,x_0-3h)+g(t_0-h,x_0-4h)}{4h^2},
\nonumber\\
&&\\
\frac{\partial^2 g_L}{\partial t^2}[t_0,x_0-h]=
&&\frac{g(t_0-h,x_0-4h)-4g(t_0,x_0-h)+g(t_0-h,x_0)}{4h^2}\nonumber\\
&&\frac{-4g(t_0,x_0-3h)+4g(t_0+h,x_0-2h)+2g(t_0-h,x_0-2h)}{4h^2},
\end{eqnarray}

and for the integration on the right

\begin{eqnarray}\label{ii-right-deriv}
\frac{\partial g_R}{\partial x}[t_0,x_0+h]=
&&\frac{-2g(t_0,x_0+3h)+2g(t_0,x_0+h)-2g(t_0+h,x_0+2h)}{4h}\nonumber\\
&&+\frac{g(t_0+h,x_0)+g(t_0+h,x_0+4h)}{4h},
\\
\frac{\partial g_R}{\partial t}[t_0,x_0+h]=
&&\frac{g(t_0+h,x_0)-2g(t_0,x_0+h)-2g(t_0-h,x_0+2h)}{4h}\nonumber\\
&&+\frac{2g(t_0+h,x_0+2h)-g(t_0+h,x_0+4h)+2g(t_0,x_0+3h)}{4h},
\\
\frac{\partial^2 g_R}{\partial x^2}[t_0,x_0+h]=
&&\frac{-2g(t_0+h,x_0+2h)+g(t_0+h,x_0)+g(t_0+h,x_0+4h)}{4h^2},
\\
\frac{\partial^2 g_R}{\partial t\partial x}[t_0,x_0+h]=
&&\frac{-g(t_0+h,x_0+4h)-2g(t_0,x_0+h)+2g(t_0,x_0+3h)+g(t_0+h,x_0)}{4h^2},
\nonumber\\
&&\\
\frac{\partial^2 g_R}{\partial t^2}[t_0,x_0+h]=
&&\frac{4g(t_0-h,x_0+2h)-4g(t_0,x_0+h)+2g(t_0+h,x_0+2h)}{4h^2}\nonumber\\
&&+\frac{g(t_0+h,x_0)+g(t_0+h,x_0+4h)-4g(t_0,x_0+3h)}{4h^2}.
\end{eqnarray}

\subsection{}
\begin{figure}
\begin{center}
\includegraphics[width=5.0in]{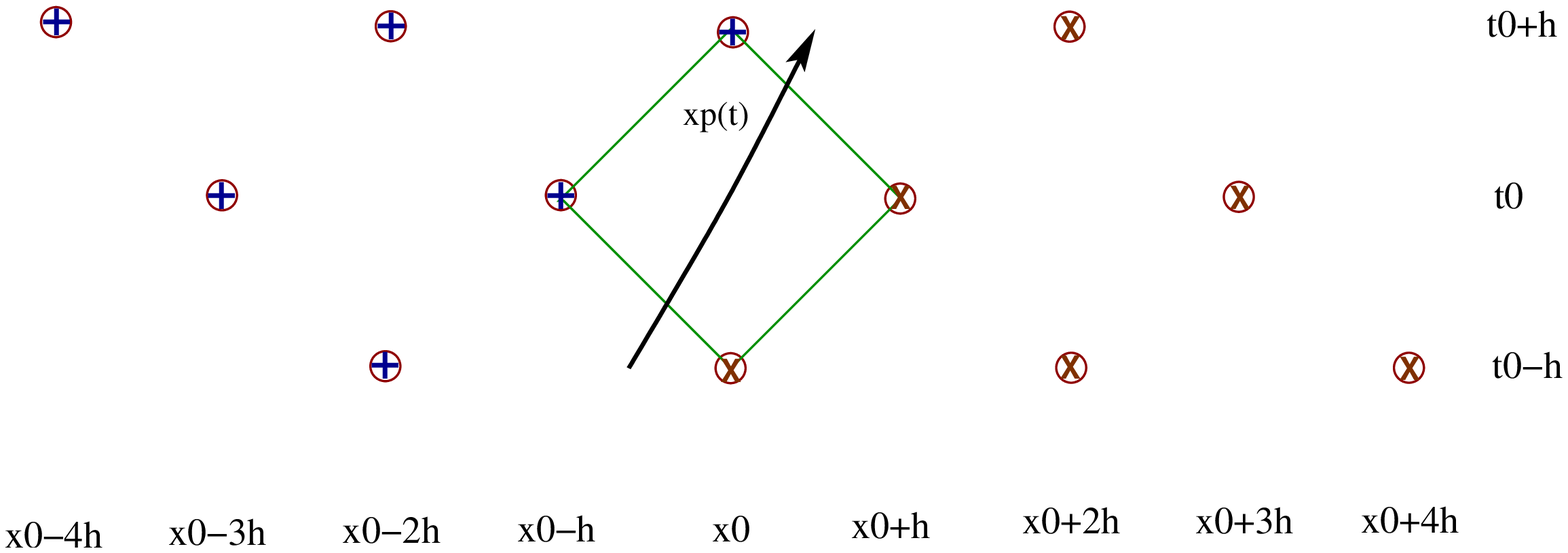}
\end{center}
\caption{Case iii) the particle enters the cell on the left and leaves it on the right. The
field $\psi$ is approximated by an expansion around the left corner
of the cell using the points labeled by $\oplus$ and around the
right corner by points labeled by $\otimes$
.}
\label{fig:TRES-LEFT}
\end{figure}

Case iii), the particle enters the cell on the left side and leaves it
on the right side. The construction of the derivatives is made taking
into account the points labeled with the crossed circle on the left
side, $\oplus$, and with the Xed-circle on the right, $\otimes$ [See
Fig.~\ref{fig:TRES-LEFT}.]  Note that if we had chosen the alternative
point $g(t_0-h,x_0-4h)$ instead of $g(t_0+h,x_0-4h)$ to the left, the
set of equations do not produce a solution.  The same applies to the
right of the particle if we had taken $g(t_0+h,x_0+4h)$ instead of
$g(t_0-h,x_0+4h)$ as the sixth grid point.

To the required order, the derivatives in terms of grid points
evaluations read explicitly for the integration on the left

\begin{eqnarray}\label{iii-left-deriv}
\frac{\partial g_L}{\partial x}[t_0,x_0-h]=
&&\frac{-2g(t_0,x_0-3h)+2g(t_0,x_0-h)-2g(t_0+h,x_0-2h)}{4h}\nonumber\\
&&+\frac{g(t_0+h,x_0)+g(t_0+h,x_0-4h)}{4h},
\\
\frac{\partial g_L}{\partial t}[t_0,x_0-h]=
&&\frac{g(t_0+h,x_0)-2g(t_0,x_0-h)-2g(t_0-h,x_0-2h)}{4h}\nonumber\\
&&+\frac{2g(t_0+h,x_0-2h)-g(t_0+h,x_0-4h)+2g(t_0,x_0-3h)}{4h},
\\
\frac{\partial^2 g_L}{\partial x^2}[t_0,x_0-h]=
&&\frac{-2g(t_0+h,x_0-2h)+g(t_0+h,x_0)+g(t_0+h,x_0-4h)}{4h^2},
\\
\frac{\partial^2 g_L}{\partial t\partial x}[t_0,x_0-h]=
&&\frac{-g(t_0+h,x_0-4h)-2g(t_0,x_0-h)+2g(t_0,x_0-3h)+g(t_0+h,x_0)}{4h^2},\nonumber\\&&
\\
\frac{\partial^2 g_L}{\partial t^2}[t_0,x_0-h]=
&&\frac{4g(t_0-h,x_0-2h)-4g(t_0,x_0-h)+2g(t_0+h,x_0-2h)}{4h^2},\nonumber\\
&&+\frac{g(t_0+h,x_0)+g(t_0+h,x_0-4h)-4g(t_0,x_0-3h)}{4h^2}
\end{eqnarray}

and for the integration on the right

\begin{eqnarray}\label{iii-right-deriv}
\frac{\partial g_R}{\partial x}[t_0,x_0+h]=
&&\frac{-2g(t_0,x_0+3h)+2g(t_0,x_0+h)-2g(t_0-h,x_0+2h)}{4h}\nonumber\\
&&+\frac{g(t_0-h,x_0)+g(t_0-h,x_0+4h)}{4h},
\\
\frac{\partial g_R}{\partial t}[t_0,x_0+h]=
&&\frac{-2g(t_0-h,x_0+2h)-g(t_0-h,x_0)+2g(t_0,x_0+h)-2g(t_0,x_0+3h)}{4h}\nonumber\\
&&+\frac{g(t_0-h,x_0+4h)+2g(t_0+h,x_0+2h)}{4h},
\\
\frac{\partial^2 g_R}{\partial x^2}[t_0,x_0+h]=
&&\frac{-2g(t_0-h,x_0+2h)+g(t_0-h,x_0)+g(t_0-h,x_0+4h)}{4h^2},
\\
\frac{\partial^2 g_R}{\partial t\partial x}[t_0,x_0+h]=
&&\frac{-g(t_0-h,x_0)+2g(t_0,x_0+h)-2g(t_0,x_0+3h)+g(t_0-h,x_0+4h)}{4h^2},\nonumber\\&&
\\
\frac{\partial^2 g_R}{\partial t^2}[t_0,x_0+h]=
&&\frac{g(t_0-h,x_0+4h)-4g(t_0,x_0+h)+g(t_0-h,x_0)-4g(t_0,x_0+3h)}{4h^2}\nonumber\\
&&+\frac{4g(t_0+h,x_0+2h)+2g(t_0-h,x_0+2h)}{4h^2}.
\end{eqnarray}

\subsection{}
\begin{figure}
\begin{center}
\includegraphics[width=5.0in]{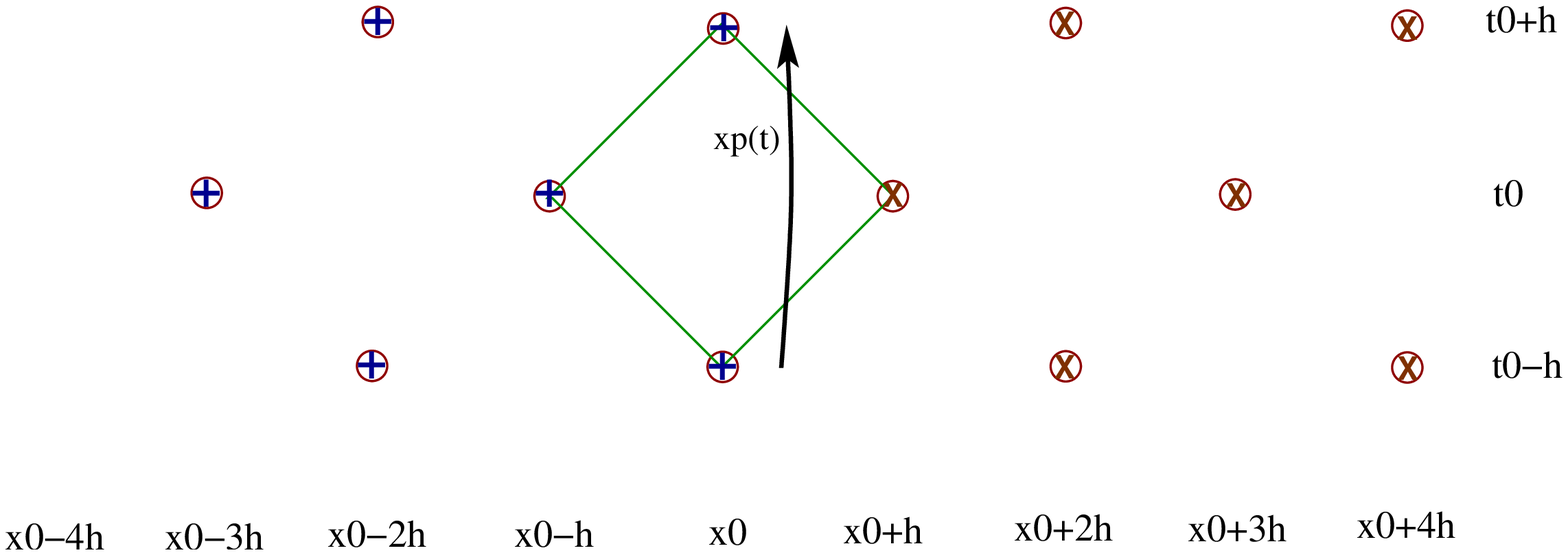}
\end{center}
\caption{Case iv) the particle enters and leaves the cell on the right. The
field $\psi$ is approximated by an expansion around the left corner
of the cell using the points labeled by $\oplus$ and around the
right corner by points labeled by $\otimes$.}
\label{fig:CUATRO-LEFT}
\end{figure}

Case iv), the particle enters and leaves the cell on the right side. The
construction of the derivatives is made taking into account the points
labeled with the crossed circle on the left side, $\oplus$, and with
the Xed-circle on the right, $\otimes$ [See Fig.~\ref{fig:CUATRO-LEFT}.]  To
the required order, the derivatives in terms of grid points
evaluations read explicitly

For the integration on the left

\begin{eqnarray}\label{iv-left-deriv}
\frac{\partial g_L}{\partial x}[t_0,x_0-h]=
&&\frac{-g(t_0+h,x_0-2h)+g(t_0-h,x_0)+g(t_0+h,x_0)-g(t_0-h,x_0-2h)}{4h},\nonumber\\&&
\\
\frac{\partial g_L}{\partial t}[t_0,x_0-h]=
&&
\frac{g(t_0+h,x_0)-g(t_0-h,x_0-2h)-g(t_0-h,x_0)+g(t_0+h,x_0-2h)}{4h},\nonumber\\&&
\\
\frac{\partial^2 g_L}{\partial x^2}[t_0,x_0-h]=
&&\frac{-g(t_0+h,x_0-2h)-2g(t_0,x_0-h)+g(t_0-h,x_0)}{4h^2}\nonumber\\
&&+\frac{g(t_0+h,x_0)-g(t_0-h,x_0-2h)+2g(t_0,x_0-3h)}{4h^2},
\\
\frac{\partial^2 g_L}{\partial t\partial x}[t_0,x_0-h]=
&&\frac{-g(t_0-h,x_0)+g(t_0+h,x_0)-g(t_0+h,x_0-2h)+g(t_0-h,x_0-2h)}{4h^2},\nonumber\\&&
\\
\frac{\partial^2 g_L}{\partial t^2}[t_0,x_0-h]=
&&\frac{3g(t_0-h,x_0-2h)-6g(t_0,x_0-h)-2g(t_0,x_0-3h)}{4h^2}\nonumber\\
&&+\frac{g(t_0-h,x_0)+g(t_0+h,x_0)+3g(t_0+h,x_0-2h)}{4h^2},
\end{eqnarray}

and for the integration on the right

\begin{eqnarray}\label{iv-right-deriv}
\frac{\partial g_R}{\partial x}[t_0,x_0+h]=
&&\frac{4g(t_0,x_0+h)+g(t_0+h,x_0+4h)+g(t_0-h,x_0+4h)}{4h}\nonumber\\
&&-\frac{4g(t_0,x_0+3h)+g(t_0-h,x_0+2h)+g(t_0+h,x_0+2h)}{4h},\nonumber\\&&
\\
\frac{\partial g_R}{\partial t}[t_0,x_0+h]=
&&\frac{3g(t_0+h,x_0+2h)-3g(t_0-h,x_0+2h)}{4h}\nonumber\\
&&\frac{-g(t_0+h,x_0+4h)+g(t_0-h,x_0+4h)}{4h},
\\
\frac{\partial^2 g_R}{\partial x^2}[t_0,x_0+h]=
&&\frac{g(t_0+h,x_0+4h)+2g(t_0,x_0+h)+g(t_0-h,x_0+4h)}{4h^2}\nonumber\\
&&-\frac{g(t_0-h,x_0+2h)+g(t_0+h,x_0+2h)+2g(t_0,x_0+3h)}{4h^2},\nonumber\\&&
\\
\frac{\partial^2 g_R}{\partial t\partial x}[t_0,x_0+h]=
&&\frac{g(t_0-h,x_0+4h)-g(t_0-h,x_0+2h)}{4h^2}\nonumber\\
&&\frac{-g(t_0+h,x_0+4h)+g(t_0+h,x_0+2h)}{4h^2},
\\
\frac{\partial^2 g_R}{\partial t^2}[t_0,x_0+h]=
&&\frac{3g(t_0-h,x_0+2h)-2g(t_0,x_0+h)}{4h^2}\nonumber\\
&&\frac{+g(t_0+h,x_0+4h)+g(t_0+h,x_0+2h)}{4h^2}\nonumber\\
&&\frac{+g(t_0-h,x_0+4h)+3g(t_0+h,x_0+2h)-6g(t_0,x_0+3h)}{4h^2}.\nonumber\\
\end{eqnarray}

\bigskip\bigskip
\noindent
{\bf References}
\bigskip
\bibliography{4torden-iop}

\end{document}